\documentclass[aps,prx,onecolumn,superscriptaddress,nofootinbib,notitlepage]{revtex4-2}
\usepackage[utf8]{inputenc}
\usepackage[T1]{fontenc}
\usepackage{lmodern}
\usepackage{amsmath,amssymb,bm,mathtools,braket}
\usepackage{graphicx}
\usepackage{booktabs}
\usepackage{xcolor}
\usepackage[colorlinks=true,linkcolor=blue,citecolor=blue,urlcolor=blue]{hyperref}

\begin{document}

\title{Resolving nonlocal state reduction via induced coherence in frequency-comb biphoton sources}

\author{Tai Hyun Yoon}
\email{thyoon@korea.ac.kr}
\affiliation{Department of Physics, Korea University, Seoul 02841, Republic of Korea}
\affiliation{Center for Molecular Spectroscopy and Dynamics, Institute for Basic Science (IBS), Seoul 02841, Republic of Korea}

\date{\today}

\begin{abstract}
How macroscopic definite outcomes and nonlocal correlations emerge from entangled superpositions across spacelike-separated regions without superluminal action or \textit{ad hoc} wavefunction collapse remains an open question in quantum foundations. Here, we show that this paradox dissolves within \textit{linear} relativistic quantum electrodynamics through an open-system, measurement-defined framework realized in two spacelike-separated nonlinear crystals driven synchronously by an optical frequency comb and seeded by a narrow-linewidth continuous-wave laser. Expanding the transverse vector potential in single-photon frequency-comb supermodes, we show that tracing out the unmeasured continuous-wave idler field yields an exact spatial dephasing Lindblad master equation whose decoherence rate is governed by the non-orthogonal state overlap of single-photon-added coherent states. Connecting this open-system map to free-vector-space quotient measurement theory, we show that single-particle interference arises from the projection of the composite state onto detector-defined conditional cosets rather than from remote collapse, establishing the operational equivalence of pump, seed, and signal phase scans. The resulting wave-particle-entanglement triality equality $P^2 + E^2 + V^2 = 1$ is saturated across the complete parameter space, while the atomic-clock-referenced comb structure embeds the signal supermode as a finite-energy codeword of a time-frequency Gottesman-Kitaev-Preskill (TF-GKP) qubit, whose stabilizer lattice $(2\Delta T,\,\omega_r)$ and logical translations $(\Delta T,\,\omega_r/2)$ are fixed by the comb repetition rate and whose conditional single-photon state is steered over the full logical Bloch sphere by a seed component shifted by half the repetition frequency. This unifies induced coherence, Bohr's complementarity, and quotient measurement theory, providing a microscopic pathway toward displacement-tolerant bosonic encoding for photonic quantum information.
\end{abstract}

\maketitle

\begin{figure}[t]
    \centering
    \includegraphics[width=\textwidth]{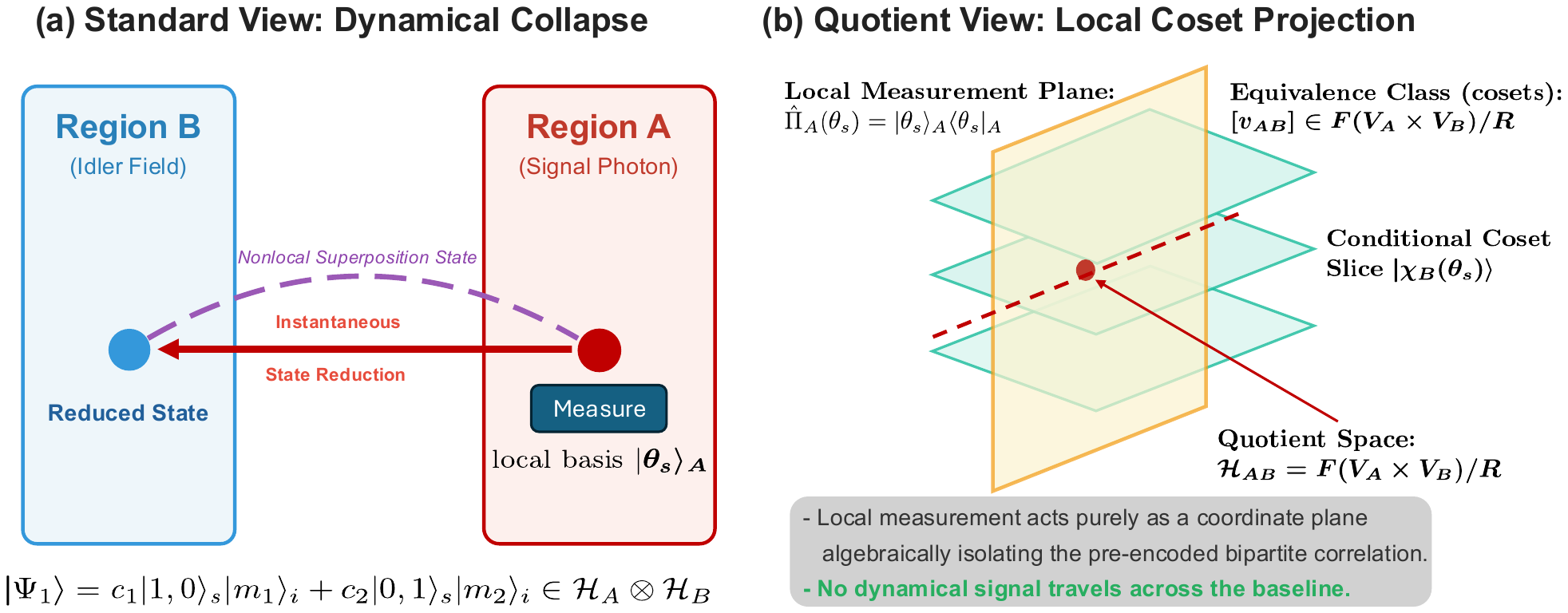}
    \caption{\textbf{Conceptual comparison of nonlocal measurement formalisms.} 
\textbf{(a) Standard Dynamical Collapse:} The conventional projection postulate treats the composite biphoton state as a rigid geometric object bridging spacelike-separated domains. Measuring the signal photon in Region A (with local basis $|\theta_s\rangle$) induces an instantaneous, non-unitary physical state reduction across the macroscopic baseline to the idler field in Region B. 
\textbf{(b) Quotient View: Local Coset Projection:} In the quotient space framework $\mathcal{H}_{AB} = F(V_A \times V_B)/R$ (detailed in the sections \emph{Foundational Questions in Subsystem Measurements and Nonlocality} and \emph{Free-Vector-Space Quotient Construction and Measurement-Defined Projection}), the composite state exists as an invariant equivalence class manifold $[v_{AB}]$ (horizontal green planes). The local measurement apparatus acts as a transversal slicing plane $\hat{\Pi}_A(\theta_s) = |\theta_s\rangle_A \langle\theta_s|_A$ within the local carrier algebra (vertical orange plane). Evaluating this local projection algebraically isolates the pre-encoded bipartite correlation, naturally revealing the conditional coset slice $|\chi_B(\theta_s)\rangle$ representing the idler field along the intersection line (dashed red line and point). Because the correlation is governed by geometric intersection rather than physical transmission, definite outcomes emerge without any dynamical signal traveling across the baseline.}
    \label{fig:fig3}
\end{figure}

Bohr's principle of complementarity \cite{Bohr1928} and the Einstein-Podolsky-Rosen (EPR) paradox \cite{Einstein1935} have stood at the core of foundational debates in quantum mechanics for nearly a century \cite{Dirac1958,Brunner2014}. Central to both puzzles is the nature of quantum measurement: why composite entangled systems, whose spatial subsystems are distributed across spacelike-separated domains, yield definite, correlated classical measurement outcomes when interrogated by local detectors [as contrasted between standard dynamical collapse in Fig.~\ref{fig:fig3}(a) and algebraic coset projection in Fig.~\ref{fig:fig3}(b)] \cite{Aspect1988,Wiseman2010,Scholes2026,Scholes2026a}. Standard quantum theory resolves this tension via the projection postulate, asserting that measurement precipitates an instantaneous state reduction across the entire spatial slice [Fig.~\ref{fig:fig3}(a)] \cite{vonNeumann1931,Wiseman2010}. However, because unitary evolution under the Schr\"odinger equation cannot transform a pure entangled state into a definite macroscopic readout without invoking non-linear modifications or subjective collapse mechanisms, the microscopic mechanism underlying state reduction remains a major open problem in modern physics \cite{Zurek2003,Wiseman2010,Scholes2026,Scholes2026a}.

In parallel, the wave-particle duality of a single quanton traversing an interferometer has been elevated from a qualitative postulate to an exact quantitative relation \cite{Greenberger1988,Jaeger1995,Englert1996,Yoon2021}. Early formulations established that path predictability $P$ and fringe visibility $V$ obey the duality inequality $P^2 + V^2 \le 1$, reaching equality only for pure quanton states \cite{Greenberger1988,Jaeger1995,Englert1996}. When the quanton becomes entangled with external degrees of freedom---such as which-path markers, unobserved environmental reservoirs, or internal spin states---the source purity $\mu_s$ degrades, leading to generalized triality relations of the form $P^2 + E^2 + V^2 = 1$, where $E$ quantifies bipartite entanglement \cite{Jakob2010,Qian2020,Yoon2021}. While theoretically profound, realizing an experimental architecture that permits continuous, independent calibration of $P$, $E$, and $V$ over their complete physical domains while rigorously enforcing relativistic microcausality has remained challenging \cite{Yoon2021}.

A distinct and powerful platform for addressing these questions is provided by nonlinear biphoton interferometry based on induced coherence without induced emission \cite{Zou1991,Wang1991,Ou1990,Heuer2015}. In this paradigm, pairs of signal and idler photons are generated in two separate nonlinear crystals \cite{Zou1991,Wang1991}. If the conjugate idler modes are aligned or seeded such that their quantum states become indistinguishable, first-order interference of the signal photons emerges without requiring heralded idler detection \cite{Zou1991,Wang1991,Lee2017,Lee2018}. Recently, it was demonstrated that dual periodically poled lithium niobate (PPLN) crystals pumped synchronously by a phase-locked optical frequency comb (OFC) and stimulated by an ultra-narrow continuous-wave (cw) seed laser can generate highly coherent single-photon frequency-comb (SPFC) supermodes \cite{Lee2018}. Because the crystals are separated by a macroscopic baseline ($L_0 \approx 1\text{ m}$) in the laboratory, the local dielectric interactions occur in spacelike-separated regions, opening a realistic avenue for exploring relativistic quantum electrodynamics (QED) and open-system measurement physics in multi-mode optical fields \cite{Lee2018,Yoon2021,Brunner2014}.

More recently, two fundamental theoretical developments have reshaped our understanding of this architecture. First, Scholes \cite{Scholes2026} demonstrated that the apparent nonlocality and collapse of entangled states can be reformulated without superluminal action by analyzing the quotient structure of tensor product spaces: an entangled state represents an equivalence class in a free vector space whose coset representatives are naturally resolved by local measurement projections. Second, Yoon \cite{Yoon2026b} showed that single-particle interference is governed not by path-accumulated delays alone, but by the relative phase between the prepared quantum state and the detector-selected basis---yielding a universal three-scan phase equivalence across pump, seed, and signal fields. Concurrently, the multi-frequency comb structure of such SPFC supermodes has been identified by Yoon \cite{Yoon2026a} as a hardware-native substrate for continuous-variable time-frequency Gottesman-Kitaev-Preskill (TF-GKP) bosonic error-correcting codes, anchoring the GKP stabilizer lattice and the logical Pauli translations directly to the underlying atomic frequency reference \cite{Gottesman2001,Fabre2020,Descamps2024,Yoon2026a}.

In this paper, we develop a relativistic QED and open-quantum-system theory of induced coherence in dual-crystal frequency-comb biphoton sources \cite{Lee2018}. Leading with the foundational dilemmas of quantum measurement, we show how this physical architecture accounts for definite, correlated local outcomes [contrasting the conventional dynamical collapse of Fig.~\ref{fig:fig3}(a) with the quotient-space coset projection of Fig.~\ref{fig:fig3}(b)] without invoking non-linear modifications or superluminal collapse. We then formulate the light-matter interaction in the transverse gauge with strict relativistic microcausality, derive the multi-mode SPFC supermode state, and show that tracing out the unmeasured idler continuum generates an exact spatial Lindblad master equation along the crystal propagation axis. We show that the effective spatial dephasing rate is fixed analytically by the non-orthogonal overlap fidelity of single-photon-added coherent states. Connecting this open-system master equation to the free-vector-space quotient formulation \cite{Scholes2026,Scholes2026a}, we demonstrate that single-particle interference is an operational projection onto a detector-defined coset slice, resolving the apparent paradox of nonlocal state reduction without superluminal signaling while saturating the quantitative triality equality $P^2 + E^2 + V^2 = 1$ \cite{Yoon2021,Yoon2026b}. Finally, we elucidate how this architecture encodes hardware-native finite-energy TF-GKP logical states---the codeword $|\bar{0}\rangle$ for a single seed frequency and the full logical Bloch sphere of the conditional single-photon state for a seed component shifted by $f_r/2$---with a stabilizer cell of symplectic area $4\pi$ and a correctable displacement region fixed by the comb \cite{Yoon2026a}, and we delineate what the seeded source can and cannot supply, outlining its potential for photonic quantum computing with displacement-tolerant logical qubits, high-dimensional quantum communication, and quantum sensing with undetected photons.

\section{Measurement of Quantum States and Nonlocal Correlations}
\label{sec:measurements_and_nonlocality}

\subsection{Foundational Questions in Subsystem Measurements and Nonlocality}
\label{subsec:foundational_questions}

In the standard formulation of quantum theory, the measurement postulate asserts that measuring an observable $\hat{A} = \sum_n a_n |a_n\rangle\langle a_n|$ on a state $|\Psi\rangle$ yields a discrete eigenvalue $a_n$ with probability $P(a_n) = |\langle a_n|\Psi\rangle|^2$, accompanied by an abrupt state reduction [contrasted with the dynamical collapse scenario in Fig.~\ref{fig:fig3}(a)] into the corresponding eigenspace \cite{vonNeumann1931,Wiseman2010}. When applied to composite entangled systems $\mathcal{H}_{AB} = \mathcal{H}_A \otimes \mathcal{H}_B$ whose subsystems are separated across spacelike intervals, this postulate introduces two long-standing foundational tensions: first, linear unitary evolution governed by the Schr\"odinger equation cannot transform a pure entangled superposition into a single definite classical meter reading without introducing {\it ad hoc} non-linear modifications or external classical collapses \cite{Zurek2003,Wiseman2010}; second, for an entangled state such as the Bell triplet or singlet, measurements performed independently on subsystem $A$ and subsystem $B$ yield outcomes that violate Bell and CHSH inequalities up to the Tsirelson bound $2\sqrt{2}$ \cite{Aspect1988,Brunner2014}. The apparent paradox lies in explaining how local measurement events remain strictly correlated without invoking non-local superluminal interactions or local hidden variables \cite{Aspect1988,Brunner2014,Yoon2026b}.

Recently, Scholes \cite{Scholes2026} proposed a resolution to these questions by analyzing measurements on separated subsystems through the algebraic structure of tensor products [illustrated via local coset projection in Fig.~\ref{fig:fig3}(b)]. Rather than viewing the composite state purely within the quotient space $\mathcal{H}_A \otimes \mathcal{H}_B$, one examines its pre-images in the free vector space $F(\mathcal{H}_A \times \mathcal{H}_B)$ [Appendix~\ref{app:quotient}] \cite{Scholes2026,Scholes2026a}. In this framework, an entangled state vector represents an entire equivalence class (coset) of formal pairs that are partitioned into distinct contextual phase classes \cite{Yoon2026b,Scholes2026,Scholes2026a}. Definite outcomes emerge because the local projection associated with the measurement apparatus isolates a specific coset representative, causing the state of the separated subsystem to simplify through local self-interference into a definite eigenstate \cite{Yoon2026b,Scholes2026,Scholes2026a}. The correlations between separated subsystems are thus pre-encoded in the global contextual phase structure of the equivalence class, precluding any superluminal dynamical action \cite{Brunner2014,Yoon2026b}.

\begin{figure}[t]
    \centering
    \includegraphics[width=\textwidth]{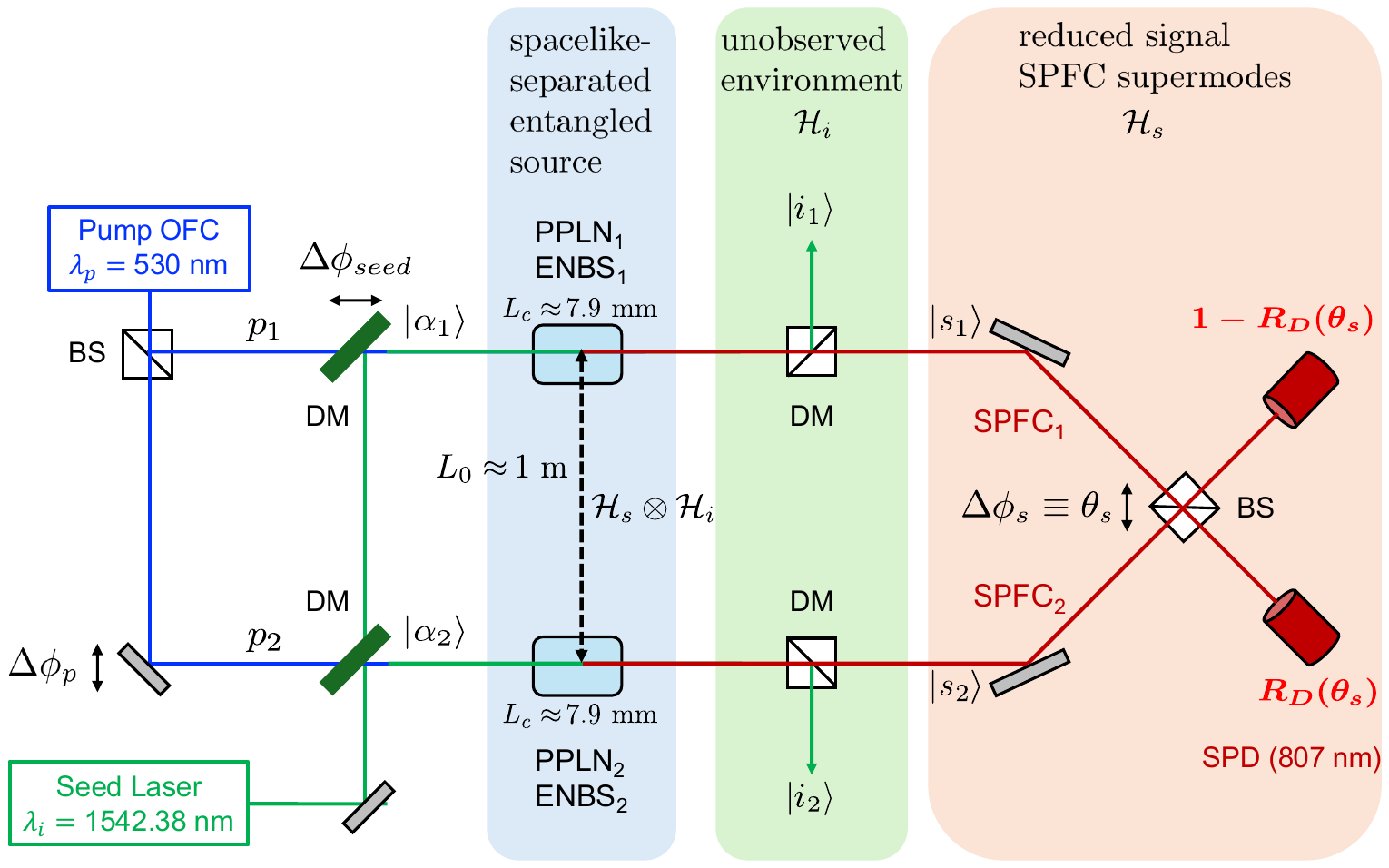}
    \caption{\textbf{Experimental realization of the quotient measurement framework in a dual-crystal single-photon frequency-comb (SPFC) architecture.} The blue region denotes the composite entangled state generated across a macroscopic spacelike-separated baseline ($L_0 \approx 1$ m) in the tensor product space $\mathcal{H}_s \otimes \mathcal{H}_i$. The green region represents the unmeasured conjugate idler field $\mathcal{H}_i$, which acts as the decoherence environment driving the exact spatial Lindblad master equation. Tracing out this continuum yields the reduced signal SPFC supermodes (red region). Here, single-particle interference is operationally resolved by projecting the state onto a detector-defined local measurement basis via the interferometric phase $\Delta\phi_s$, without invoking superluminal state reduction. The universal three-scan phase equivalence is governed by the independent manipulation of the pump ($\Delta\phi_p$), seed ($\Delta\phi_{seed}$), and signal ($\Delta\phi_s\equiv\theta_s$) phases.}
    \label{fig:fig1}
\end{figure}

\subsection{Physical Realization via the Dual-Crystal SPFC Architecture}
\label{subsec:spfc_realization}

While the quotient-space formulation provides an appealing algebraic perspective, it has largely been discussed in terms of abstract finite-dimensional spin systems \cite{Scholes2026}. Here, we demonstrate that the dual-crystal stimulated parametric down-conversion system provides a concrete, continuous-variable-anchored physical realization of this measurement theory \cite{Lee2018,Yoon2021,Yoon2026a}.
In our experimental configuration (illustrated in Fig.~\ref{fig:fig1}), the two PPLN crystals are physically separated by $L_0 \approx 1\text{ m}$ in the laboratory \cite{Lee2018,Yoon2021}. Microcausality guarantees that the local field algebras $\mathcal{A}(V_1)$ and $\mathcal{A}(V_2)$ commute \cite{Peskin1995}.

\begin{figure}[t]
    \centering
    \includegraphics[width=\textwidth]{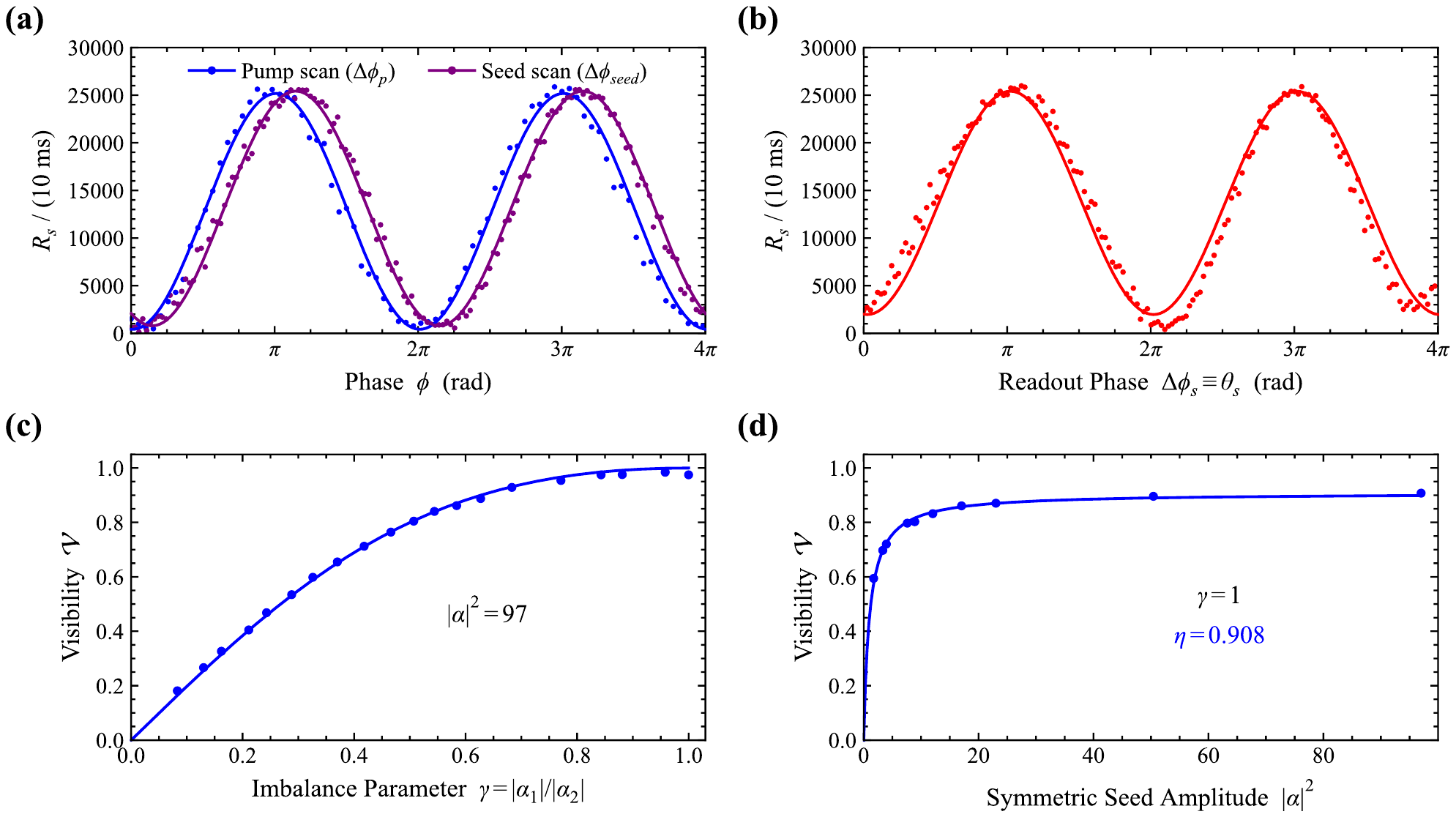}
    \caption{\textbf{Operational validation of the three-scan phase equivalence and spatial Lindblad dephasing.} Experimental data (reproduced from Ref.~\cite{Lee2018}) analyzed under the free-vector-space quotient measurement framework. \textbf{(a)} Single-photon counting rates $R_s$ exhibiting near-unit visibility interference upon independent modulation of the macroscopic source preparation phases: the pump comb ($\Delta\phi_p$) and the continuous-wave idler seed ($\Delta\phi_{seed}$). \textbf{(b)} Identical interference statistics generated by rotating the local measurement basis parameterized by the readout phase $\Delta\phi_s \equiv \theta_s$ at the signal detector. The unified phase domain ($0 \to 4\pi$) across panels (a) and (b) empirically validates the universal scalar invariant $\Delta\Phi = \theta_s - \Delta\phi_p + \Delta\phi_{seed}$, confirming that interference emerges strictly from local coset projections without requiring superluminal state reduction. \textbf{(c)} and \textbf{(d)} Environment-induced decoherence mapped via the interference visibility $\mathcal{V}$ (blue data points, corresponding to the pump configuration in panel a). The experimental data are plotted as a function of the local conjugate seed parameters: the balance ratio $\gamma$ (at macroscopic seed $|\alpha|^2 \approx 97$) and the symmetric seed photon number $|\alpha|^2$ (at $\gamma = 1$). The theoretical curve in (d) includes an empirical saturation factor $\eta = 0.908$ to account for non-ideal spatial mode overlap. Solid curves represent the exact analytical bounds derived from the spatial Lindblad master equation, empirically demonstrating the continuous degradation of source purity as which-path distinguishability is partitioned to the unobserved idler reservoir.}
    \label{fig:fig2}
\end{figure}

In the abstract quotient model, the contextual phase classes arise from the formal association of the superposition phase with either the first or second factor of a tensor product pair \cite{Yoon2026b,Scholes2026}. In our dual-crystal SPFC platform (see Fig.~\ref{fig:fig1}), these abstract contextual phases correspond directly to experimentally calibrated, physical optical path phases \cite{Lee2018,Yoon2021}:
\begin{equation}
\Delta\Phi = \Delta\phi_s - \Delta\phi_p + \Delta\phi_{\text{seed}},
\label{eq:total_phase_secB}
\end{equation}
where $\Delta\phi_p$ is the relative pump comb phase, $\Delta\phi_{\text{seed}}$ is the narrow-linewidth idler seed phase, and $\Delta\phi_s$ is the signal interferometric phase \cite{Lee2018,Yoon2026b}. As established in Appendix~\ref{app:measurement}, the detector projection $\hat{\Pi}_A(\theta_s) = |\theta_s\rangle_A \langle \theta_s|_A$ does not collapse the distant idler beam through any instantaneous physical action \cite{Yoon2021,Yoon2026b}. Instead, the linear contraction of the dual bra vector ${}_A\langle\theta_s| = \frac{1}{\sqrt{2}}({}_s\langle 1,0| + e^{-i\theta_s}{}_s\langle 0,1|)$ extracts the conditional coset slice:
\begin{equation}
|\chi_B(\theta_s)\rangle = \frac{1}{\sqrt{2}} \left( c_1 |m_1\rangle_i + c_2 e^{-i\theta_s} |m_2\rangle_i \right).
\label{eq:chi_B_secB}
\end{equation}
The single-photon count rate $R_D(\theta_s) \propto \langle\chi_B(\theta_s)|\chi_B(\theta_s)\rangle$ is the operational manifestation of self-interference across the coset slice, producing identical sinusoidal fringe visibility $V \approx 0.99$ whether $\Delta\phi_p$, $\Delta\phi_{\text{seed}}$, or $\Delta\phi_s\equiv\theta_s$ is varied [as demonstrated in the experimental scans of Figs.~\ref{fig:fig2}(a) and \ref{fig:fig2}(b)] \cite{Lee2018,Yoon2026b}. This validates the foundational concept demonstrated by Yoon \cite{Yoon2026b} that single-particle interference is governed by the relative phase between the prepared quantum state and the detector-defined measurement basis.

Crucially, our system extends the abstract measurement framework beyond isolated pure states by incorporating open quantum system dynamics \cite{Lindblad1976,Carmichael1993}. In realistic measurements, the conjugate subsystem is often unmeasured \cite{Lee2018,Yoon2021}. Tracing out this reservoir generates an exact spatial dephasing Lindblad dissipator with a dephasing rate $\Gamma_{\text{deph}} = -(1/L_c)\ln F$ along the propagation axis over the crystal interaction length $L_c$ [Appendix~\ref{app:lindblad}] \cite{Yoon2021}. This links the algebraic quotient projection directly to the quantum state overlap $F = |\langle m_2|m_1\rangle_i|$: when $F \to 1$ (macroscopic coherent seed, $|\alpha|^2 \gg 1$), the distinguishability vanishes ($E \to 0$), the dephasing rate drops to zero ($\Gamma_{\text{deph}} \to 0$), and the system reproduces the ideal, dissipation-free quotient projection with unit visibility ($V \to 1$) \cite{Lee2018,Yoon2021}; conversely, when $F \to 0$ (spontaneous vacuum seed, $|\alpha| \to 0$), the environment acquires which-path information ($E \to 1$), driving $\Gamma_{\text{deph}} \to \infty$ and reducing the signal state to a classical statistical mixture without interference ($V \to 0$) \cite{Lee2018,Yoon2021}. Thus, our dual-source biphoton model [as experimentally verified by the measured dephasing and visibility trends in Fig.~\ref{fig:fig2}(d)] bridges the gap between the algebraic measurement formulation of separated subsystems and physical open-system decoherence \cite{Zurek2003,Yoon2021,Yoon2026b}.

\subsection{Conceptual Synthesis: State Reduction versus Quotient Projection}
\label{subsec:classification_issues}

The realization of single-particle interference in our dual-crystal SPFC architecture provides a clear conceptual contrast between standard dynamical collapse [Fig.~\ref{fig:fig3}(a)] and algebraic quotient projection [Fig.~\ref{fig:fig3}(b)].

Under the conventional projection postulate [Fig.~\ref{fig:fig3}(a)], measurement is regarded as an active, non-unitary operation that dynamically alters the physical state across spacelike-separated regions \cite{vonNeumann1931,Wiseman2010}. In contrast, within the quotient framework $F(V_A \times V_B)/R$ [Fig.~\ref{fig:fig3}(b), see Appendix~\ref{app:quotient}], the prepared entangled state $[v_{AB}]$ represents an invariant equivalence class encoding all possible bilinear contractions prior to measurement \cite{Yoon2026b,Scholes2026}. When an observer at station $A$ measures the signal field with projector $\hat{\Pi}_A(\theta_s)$, this operation does not transmit a physical signal across the laboratory baseline $L_0 \approx 1\text{ m}$ to collapse the idler beam \cite{Lee2018,Brunner2014}. Instead, the local projector acts as a coordinate basis selector within the local carrier space $V_A$, isolating the corresponding coset slice $|\chi_B(\theta_s)\rangle$ in $V_B$ through linear contraction [Eq.~(\ref{eq:chi_B_secB})] \cite{Yoon2026b,Scholes2026}. Because the two local projections $\hat{\Pi}_A \otimes \hat{I}_B$ and $\hat{I}_A \otimes \hat{\Pi}_B$ commute, the joint statistics of a measurement on the signal and a measurement (or herald) on the idler are independent of the order in which the projections are applied, and a classical record may combine the two outcomes in either order; causal ordering is required only where a later operation is physically conditioned on an earlier outcome. Because state reduction is reinterpreted as an algebraic coset selection rather than a physical wave-packet collapse, strict relativistic causality is naturally preserved; physical observables defined by the transverse electromagnetic vector potential operators $\hat{\mathbf{A}}_\perp(x_1)$ and $\hat{\mathbf{A}}_\perp(x_2)$ at distinct spacetime points satisfy the exact commutation relation $[\hat{\mathbf{A}}_\perp(x_1), \hat{\mathbf{A}}_\perp(x_2)] = 0$ across all spacelike intervals [Appendix~\ref{app:qed}] \cite{Peskin1995}.

Furthermore, this algebraic perspective clarifies the physical nature of interferometric phases \cite{Yoon2026b}. In standard descriptions, changing an optical path delay in an interferometer is treated as a local dynamical phase accumulation, whereas changing a state preparation parameter is considered an intrinsic state modification \cite{Lee2018,Yoon2026b}. The quotient formulation demonstrates that these operations are mathematically equivalent: shifting the pump comb phase $\Delta\phi_p$, the seed phase $\Delta\phi_{\text{seed}}$, or the detector projection angle $\theta_s = \Delta\phi_s$ corresponds to equivalent parameterizations of the same scalar invariant $\Delta\Phi = \Delta\phi_s - \Delta\phi_p + \Delta\phi_{\text{seed}}$ governing the inner product of the coset slice \cite{Lee2018,Yoon2026b}.

Finally, combining quotient projections with open-system dynamics resolves how environmental distinguishability degrades interference \cite{Yoon2021,Yoon2026b}. When the idler field is unobserved, tracing over $V_B$ does not imply that the signal photon randomly collapsed at the source; rather, the loss of interference visibility is determined by the non-orthogonal overlap $F$ of the idler environment states, which directly fixes the dephasing rate $\Gamma_{\text{deph}} = -(1/L_c)\ln F$ and saturates the triality relation $P^2 + E^2 + V^2 = 1$ \cite{Jakob2010,Yoon2021}. Having established these conceptual foundations, we now turn to the complete mathematical formulation within relativistic QED.

\section{Theoretical Framework}
\label{sec:theory}

\subsection{Microcausality and Canonical Quantization in Spacelike-Separated Dual PPLN Volumes}
\label{subsec:quantization}

We consider two identical periodically poled lithium niobate ($\mathrm{LiNbO}_3$, PPLN) crystals, denoted $\mathrm{NL}_1$ and $\mathrm{NL}_2$, situated in the laboratory with spatial separation $L_0 = |\mathbf{r}_2 - \mathbf{r}_1| \approx 1\text{ m}$ [see experimental layout in Fig.~\ref{fig:fig1}] \cite{Lee2018,Yoon2021}. In the transverse (Coulomb) gauge $\boldsymbol{\nabla}\cdot\hat{\mathbf{A}}(\mathbf{r},t) = 0$, the quantized relativistic electromagnetic vector potential operator in the interaction picture is expanded in terms of continuum plane-wave Fock modes \cite{Lee2018}:
\begin{equation}
\begin{split}
\hat{\mathbf{A}}_\perp(\mathbf{r}, t) &= \sum_{\lambda=1,2} \int \frac{d^3 k}{(2\pi)^{3/2}} \sqrt{\frac{\hbar}{2\varepsilon_0 n^2(\omega_\mathbf{k}) \omega_\mathbf{k}}}\; \boldsymbol{\epsilon}_{\mathbf{k},\lambda} \\
&\quad\times \left[ \hat{a}_{\mathbf{k},\lambda} e^{i(\mathbf{k}\cdot\mathbf{r} - \omega_\mathbf{k} t)} + \hat{a}_{\mathbf{k},\lambda}^\dagger e^{-i(\mathbf{k}\cdot\mathbf{r} - \omega_\mathbf{k} t)} \right],
\end{split}
\label{eq:A_field}
\end{equation}
where $\boldsymbol{\epsilon}_{\mathbf{k},\lambda}$ are orthonormal transverse polarization unit vectors, $n(\omega_\mathbf{k})$ is the refractive index along the active dielectric principal axis, and the annihilation and creation operators satisfy the canonical equal-time commutation relations $[\hat{a}_{\mathbf{k},\lambda}, \hat{a}_{\mathbf{k}',\lambda'}^\dagger] = \delta_{\lambda\lambda'} \delta^3(\mathbf{k}-\mathbf{k}')$. For any two spacetime points $x_1 = (c t_1, \mathbf{r}_1)$ and $x_2 = (c t_2, \mathbf{r}_2)$ lying in spacelike configuration, $(x_1 - x_2)^2 = c^2 (t_1 - t_2)^2 - |\mathbf{r}_1 - \mathbf{r}_2|^2 < 0$, microcausality guarantees that the field commutator vanishes identically [Appendix~\ref{app:qed} for the relativistic QED and Jordan--Pauli function derivation] \cite{Peskin1995}:
\begin{equation}
\left[ \hat{\mathbf{E}}_\perp(x_1), \hat{\mathbf{E}}_\perp(x_2) \right] = 0, \quad \left[ \hat{\mathbf{A}}_\perp(x_1), \hat{\mathbf{A}}_\perp(x_2) \right] = 0.
\label{eq:microcausality}
\end{equation}
Consequently, local operations performed inside the finite interaction volume $V_1$ of crystal 1 commute with local operations inside volume $V_2$ of crystal 2 \cite{Lee2018,Yoon2021}.

For type-0 quasi-phase matching ($e \to e + e$ interaction along the extraordinary crystalline axis $\hat{\mathbf{z}}$) \cite{Lee2018,Yoon2021}, the polarization unit vectors of pump, signal, and idler fields remain parallel to $\hat{\mathbf{z}}$. Defining one-dimensional paraxial propagating wavepacket operators along spatial path $j \in \{1, 2\}$, we have \cite{Fabre2020,Brecht2015}:
\begin{equation}
\hat{a}_{\mu,j}(\omega) = \int d^2 k_\perp u_{\mu,j}^*(\mathbf{k}_\perp) \hat{a}_\mu(\mathbf{k}_\perp, k_z(\omega)),
\label{eq:mode_decomp}
\end{equation}
where $\mu \in \{s, i\}$ indexes the signal and idler fields, respectively, and $u_{\mu,j}(\mathbf{k}_\perp)$ designates the normalized transverse spatial mode profile \cite{Yoon2021}. These operators obey the one-dimensional continuum commutators $[\hat{a}_{\mu,j}(\omega), \hat{a}_{\nu,k}^\dagger(\omega')] = \delta_{\mu\nu}\delta_{jk} \delta(\omega - \omega')$.

\subsection{Multi-Mode Frequency Comb Hamiltonian and SPFC Supermode Representation}
\label{subsec:comb_hamiltonian}

The two crystals are pumped synchronously by a phase-locked optical frequency comb centered at $\lambda_p = 530\text{ nm}$ \cite{Lee2018,Yoon2021}. In the time domain, the pump pulse train is characterized by repetition period $\Delta T = 1/f_r$ and carrier-envelope offset (CEO) phase drift $\Delta\phi_{\text{ceo}}$ per round trip \cite{Lee2018,Cundiff2003,Jones2000}. At the entrance face of crystal $j$, the classical pump field is expressed as \cite{Lee2018}:
\begin{align}
E_p^{(j)}(t) &= e^{i\phi_{p,j}} \sum_{n=-\infty}^\infty A_n e^{-i\omega_{p,n} (t - t_j)}, \label{eq:pump_comb} \\
\omega_{p,n} &= 2\pi(n f_r + f_{\text{ceo}}), \quad f_{\text{ceo}} = \frac{f_r \Delta\phi_{\text{ceo}}}{2\pi},
\end{align}
where $\phi_{p,j}$ is an experimentally tunable path phase \cite{Lee2018,Yoon2021}, and the discrete Fourier amplitude $A_n$ 
associated with the Gaussian envelope of pulse duration $\delta \approx 8.4\text{ ps}$ is given by
\begin{equation}
A_n = \mathcal{E}_p \frac{\sqrt{2\pi}\delta}{\Delta T} \exp\!\left[ -\frac{n^2}{2\left(\frac{\Delta T}{2\pi\delta}\right)^{\!2}} \right].
\label{eq:Fourier_amplitude}
\end{equation}

Simultaneously, an ultra-narrow continuous-wave laser of central frequency $\omega_{\text{seed}} = 2\pi c / \lambda_i$ ($\lambda_i = 1542.38\text{ nm}$, linewidth $\Delta\nu < 10\text{ Hz}$) seeds the idler mode in both crystals \cite{Lee2018,Yoon2021}. The seed state in crystal $j$ is a coherent state $|\alpha_j\rangle = \hat{D}_j(\alpha_j)|0\rangle$, with amplitudes $\alpha_1 = \gamma |\alpha| e^{i\phi_{i,1}}$ and $\alpha_2 = |\alpha| e^{i\phi_{i,2}}$, where the variable transmission ratio $\gamma = |\alpha_1|/|\alpha_2| \in [0, 1]$ is set by the seed beam splitter \cite{Lee2018,Yoon2021}.

The field-matter interaction within the two non-overlapping dielectric media is governed by the second-order nonlinear Hamiltonian \cite{Lee2018,Ou1990,Wang1991}:
\begin{equation}
\begin{split}
\hat{H}_I(t) &= \sum_{j=1,2} \varepsilon_0 \chi^{(2)} \int_{V_j} d^3 r \\
&\quad\times \left[ E_p^{(j)+}(\mathbf{r}, t) \hat{E}_{s,j}^{(-)}(\mathbf{r}, t) \hat{E}_{i,j}^{(-)}(\mathbf{r}, t) + \text{h.c.} \right].
\end{split}
\label{eq:H_I}
\end{equation}
Substituting the modal expansions into Eq.~(\ref{eq:H_I}) yields \cite{Lee2018,Christ2013}:
\begin{equation}
\begin{split}
\hat{H}_I(t) &= \hbar \sum_{j=1,2} \sum_{n} \iint d\omega_s\, d\omega_i \, g_n(\omega_s, \omega_i)\, e^{i\phi_{p,j}} \\
&\quad\times e^{-i(\omega_{p,n} - \omega_s - \omega_i)t}\, \hat{a}_{s,j}^\dagger(\omega_s)\, \hat{a}_{i,j}^\dagger(\omega_i) + \text{h.c.},
\end{split}
\label{eq:H_I_modes}
\end{equation}
where $g_n(\omega_s, \omega_i) = g_0 A_n \operatorname{sinc}( \Delta k_n L_c / 2 ) e^{-i\Delta k_n L_c / 2}$, with crystal interaction length $L_c \approx 7.9\text{ mm}$ and phase mismatch $\Delta k_n = k_p(\omega_{p,n}) - k_s(\omega_s) - k_i(\omega_i) - 2\pi/\Lambda_0$ for grating period $\Lambda_0 \approx 7.3\ \mu\text{m}$ \cite{Lee2018,Yoon2021}.

Because the seed laser spectral distribution approximates a Dirac delta function, $\alpha(\omega_i) \approx \alpha \, \delta(\omega_i - \omega_{\text{seed}})$, time integration over interaction duration $t \gg \Delta T$ enforces strict energy conservation at each comb tooth: $\omega_{s,n} = \omega_{p,n} - \omega_{\text{seed}} = 2\pi(n f_r + f_{\text{ceo}}) - \omega_{\text{seed}}$ \cite{Lee2018}. We define the single-photon frequency comb (SPFC) supermode creation operator $\hat{S}_j^\dagger$ on arm $j$ as the normalized discrete coherent sum \cite{Lee2018,Yoon2026a}:
\begin{equation}
\hat{S}_j^\dagger \equiv \sum_{n} \frac{A_n \operatorname{sinc}(\Delta k_n L_c / 2)}{\sqrt{\sum_m |A_m \operatorname{sinc}(\Delta k_m L_c / 2)|^2}}\,\hat{a}_{s,j}^\dagger(\omega_{s,n}),
\label{eq:supermode_def}
\end{equation}
satisfying $[\hat{S}_j, \hat{S}_k^\dagger] = \delta_{jk}$ and $[\hat{S}_j, \hat{S}_k] = 0$. The SPFC supermode is a coherent single-photon wavepacket spanning $N \approx 209$ comb lines within the phase-matching bandwidth \cite{Lee2018,Ren2014}. The teeth of this supermode lie on the grid $\omega_{s,n} = \Omega_0 + n\omega_r$, with the fractional offset $\Omega_0 \equiv 2\pi f_{\text{ceo}} - \omega_{\text{seed}} \pmod{\omega_r}$ set jointly by the carrier-envelope offset of the pump comb and by the seed frequency. Shifting either $f_{\text{ceo}}$ or $\omega_{\text{seed}}$ by $f_r/2$ displaces the entire signal comb by half a tooth spacing; this offset degree of freedom is the one that carries the logical information in the section \emph{Hardware-Anchored Time-Frequency Stabilizers and Continuous-Variable Grid Embedding}.

\subsection{Bipartite Entangled Nonlinear Biphoton State}
\label{subsec:bipartite_state}

In the low parametric gain limit ($|g_0 t A_n| \ll 1$) \cite{Lee2018,Yoon2021,Heuer2015}, the time-evolution operator $\hat{U}(t) \approx \hat{I} - \frac{i}{\hbar}\int_0^t \hat{H}_I(t') dt'$ acts on the seeded initial state $|\Psi(0)\rangle = |0\rangle_{s_1} |0\rangle_{s_2} \otimes |\alpha_1\rangle_{i_1} |\alpha_2\rangle_{i_2}$ \cite{Lee2018}. Dropping the vacuum component via conditional single-photon registration, the normalized single-pair emission state is \cite{Lee2018,Yoon2021}:
\begin{equation}
|\Psi_1\rangle = c_1 |1,0\rangle_s |m_1\rangle_i + c_2 |0,1\rangle_s |m_2\rangle_i,
\label{eq:psi_ENBS}
\end{equation}
where $|1,0\rangle_s \equiv \hat{S}_1^\dagger |0\rangle_{s_1} |0\rangle_{s_2}$ and $|0,1\rangle_s \equiv \hat{S}_2^\dagger |0\rangle_{s_1} |0\rangle_{s_2}$ denote orthonormal path-encoded SPFC states \cite{Yoon2021}, with real normalization coefficients (the pump path phases $e^{i\phi_{p,j}}$ of Eq.~(\ref{eq:H_I_modes}) are absorbed into the relative phase of the idler states below):
\begin{equation}
c_j = \frac{\sqrt{1 + |\alpha_j|^2}}{\sqrt{2 + |\alpha_1|^2 + |\alpha_2|^2}}, \quad |c_1|^2 + |c_2|^2 = 1.
\label{eq:norm_coeff}
\end{equation}
The conjugate idler states $|m_1\rangle_i$ and $|m_2\rangle_i$ represent bipartite states containing single-photon-added coherent states (SPACS) \cite{Lee2018,Yoon2021,Agarwal1991,Zavatta2004}:
\begin{equation}
|m_1\rangle_i = |\alpha_1, 1\rangle_{i_1} |\alpha_2\rangle_{i_2}, \quad |m_2\rangle_i = |\alpha_1\rangle_{i_1} |\alpha_2, 1\rangle_{i_2},
\label{eq:m_states}
\end{equation}
where the normalized SPACS on path $j$ is generated by bosonic creation on the seed: $|\alpha_j, 1\rangle_{i_j} = \hat{a}_{i,j}^\dagger |\alpha_j\rangle_{i_j} / \sqrt{1 + |\alpha_j|^2}$ \cite{Lee2018,Yoon2021,Agarwal1991,Zavatta2004}.

\subsection{Open Quantum System Formulation: Tracing Out the Undetected Continuum}
\label{subsec:open_system}

In unheralded single-photon interferometry, the conjugate idler modes are discarded without detection \cite{Lee2018,Yoon2021,Zou1991,Wang1991}. Treating the idler Hilbert space $\mathcal{H}_i$ as an unobserved reservoir $\mathcal{R}$, the reduced density matrix of the signal SPFC quanton $\hat{\rho}_s = \operatorname{Tr}_i [ |\Psi_1\rangle \langle \Psi_1| ]$ in the orthonormal basis $\{|1,0\rangle_s, |0,1\rangle_s\}$ is \cite{Lee2018,Yoon2021}:
\begin{equation}
\hat{\rho}_s = \begin{pmatrix} \rho_{11} & \rho_{12} \\ \rho_{21} & \rho_{22} \end{pmatrix} = \begin{pmatrix} |c_1|^2 & c_1 c_2^* \langle m_2 | m_1 \rangle_i \\ c_1^* c_2 \langle m_1 | m_2 \rangle_i & |c_2|^2 \end{pmatrix}.
\label{eq:reduced_density}
\end{equation}
The off-diagonal coherence is modulated by the idler state overlap: $\langle m_2 | m_1 \rangle_i = {}_{i_1}\langle \alpha_1 | \alpha_1, 1 \rangle_{i_1} \cdot {}_{i_2}\langle \alpha_2, 1 | \alpha_2 \rangle_{i_2} = F e^{-i\Delta\theta}$, where $\Delta\theta = \Delta\phi_p - \Delta\phi_{\text{seed}}$ and the state overlap fidelity $F \equiv |\langle m_2 | m_1 \rangle_i|$ evaluates to \cite{Lee2018,Yoon2021}:
\begin{equation}
F = \frac{|\alpha_1| |\alpha_2|}{\sqrt{1 + |\alpha_1|^2} \sqrt{1 + |\alpha_2|^2}} = \frac{\gamma |\alpha|^2}{\sqrt{(1+|\alpha|^2)(1+\gamma^2|\alpha|^2)}},
\label{eq:F_def}
\end{equation}
with $|\alpha_2| = |\alpha|$ and $|\alpha_1| = \gamma |\alpha|$ as above \cite{Yoon2021}.

The non-unitary evolution induced by tracing out the idler modes is expressed as a completely positive trace-preserving map $\mathcal{E}(\hat{\rho}) = \sum_{k} \hat{M}_k \hat{\rho} \hat{M}_k^\dagger$ \cite{Kraus1983,Stinespring1955,AlickiLendi1987}. In the single-photon subspace, the Kraus operators take the canonical form \cite{Kraus1983}:
\begin{equation}
\hat{M}_0 = \begin{pmatrix} 1 & 0 \\ 0 & F \end{pmatrix}, \quad \hat{M}_1 = \begin{pmatrix} 0 & 0 \\ 0 & \sqrt{1 - F^2} \end{pmatrix},
\label{eq:kraus_ops}
\end{equation}
satisfying $\hat{M}_0^\dagger \hat{M}_0 + \hat{M}_1^\dagger \hat{M}_1 = \hat{I}$. In the laboratory reference frame, the extraordinary optical axes of the two PPLN crystals are oriented vertically, perpendicular to the optical table, along the $\hat{\mathbf{z}}$ direction, fixing the linear polarization of the interacting beams. The beams propagate along the longitudinal direction with wavevectors $\mathbf{k}_j = k \hat{\mathbf{y}}$ ($j \in \{1,2\}$), which together with the transverse horizontal axis $\hat{\mathbf{x}}$ forms a standard right-handed Cartesian coordinate system $(\hat{\mathbf{x}}, \hat{\mathbf{y}}, \hat{\mathbf{z}})$. Consequently, the periodic domain inversion grating vector $\mathbf{K}_0 = (2\pi/\Lambda_0)\hat{\mathbf{y}}$ is directed along the $\hat{\mathbf{y}}$ axis. 

Along the spatial propagation coordinate $y$, the dephasing dynamics can be cast into a spatial Lindblad master equation (see Appendix~\ref{app:lindblad} for the microscopic derivation and the connection between spatial and temporal dissipation rates) \cite{Lindblad1976,Carmichael1993}:
\begin{equation}
\frac{d\hat{\rho}_s(y)}{dy} = -\frac{i}{\hbar} \left[ \hat{H}_{\text{prop}}, \hat{\rho}_s(y) \right] + \frac{\Gamma_{\text{deph}}}{2} \left( \hat{\sigma}_z \hat{\rho}_s(y) \hat{\sigma}_z - \hat{\rho}_s(y) \right),
\label{eq:lindblad}
\end{equation}
where $\hat{\sigma}_z = |1,0\rangle_s \langle 1,0| - |0,1\rangle_s \langle 0,1|$, $\hat{H}_{\text{prop}}/\hbar$ is the paraxial propagation generator (of dimension inverse length), and the effective spatial dephasing rate (also of dimension $\mathrm{m}^{-1}$) is:
\begin{equation}
\Gamma_{\text{deph}} = -\frac{1}{L_c} \ln F.
\label{eq:dephasing_rate}
\end{equation}
Here, the spatial derivative reflects the paraxial beam propagation along the crystal interaction length $L_c \approx 7.9\text{ mm}$, where the spatial dephasing coefficient $\Gamma_{\text{deph}}$ directly determines the attenuation of off-diagonal coherence per unit propagation distance [as verified by the measured dephasing trends in Fig.~\ref{fig:fig2}(d)] \cite{Lee2018,Yoon2021}. When the seed field vanishes ($|\alpha| \to 0$), $F \to 0$ and $\Gamma_{\text{deph}} \to \infty$, causing complete decoherence of the signal quanton into a statistical mixture \cite{Lee2018,Yoon2021}. Conversely, in the macroscopic seed limit ($|\alpha|^2 \gg 1$), $F \to 1$ and $\Gamma_{\text{deph}} \to 0$, protecting the signal superposition from environment-induced dephasing \cite{Lee2018,Yoon2021,Zurek2003}.

\subsection{Free-Vector-Space Quotient Construction and Measurement-Defined Projection}
\label{subsec:quotient_measurement}

To eliminate paradoxes associated with non-local wavefunction collapse across spacelike separations, we formalize the measurement process using the quotient structure of tensor products (see Appendix~\ref{app:quotient} for algebraic details) \cite{Wiseman2010,Yoon2026b,Scholes2026}. Let $V_A = \mathcal{H}_s$ and $V_B = \mathcal{H}_i$ denote the local vector spaces associated with spacelike-separated regions. The composite Hilbert space $\mathcal{H}_{AB} \cong V_A \otimes V_B$ is rigorously defined as the quotient of the free vector space $F(V_A \times V_B)$ modulo the subspace $R$ generated by bilinear equivalences: $\mathcal{H}_{AB} = F(V_A \times V_B) / R$. Before quotient projection, the physical state vector in the free space is an equivalence class $[v_{AB}]$ represented by the formal sum:
\begin{equation}
v_{AB} = c_1 \left( |1,0\rangle_s, |m_1\rangle_i \right) + c_2 \left( |0,1\rangle_s, |m_2\rangle_i \right) \in F(V_A \times V_B).
\label{eq:free_vector}
\end{equation}

A local interferometric measurement performed at the signal detection plane combines the two arms on the symmetric signal beam splitter (BS in Fig.~\ref{fig:fig1}) followed by single-photon detection with an SPD at $807\text{ nm}$ \cite{Yoon2021}. The measurement observable is defined entirely within the local carrier algebra $\mathcal{B}(V_A)$ via the projection operator $\hat{\Pi}_A(\theta_s) = |\theta_s\rangle_A \langle \theta_s|_A$, onto the detector-selected basis state \cite{Yoon2021,Yoon2026b}:
\begin{equation}
|\theta_s\rangle_A = \frac{1}{\sqrt{2}} \left( |1,0\rangle_s + e^{i\theta_s} |0,1\rangle_s \right),
\label{eq:PVM_local}
\end{equation}
where $\theta_s = \Delta\phi_s = k_s \Delta x_s$ is the signal interferometric phase \cite{Lee2018,Yoon2021}. Applying the local operation $\hat{\Pi}_A(\theta_s) \otimes \hat{I}_B$ to the composite state $[v_{AB}]$, the projection resolves the state along the measurement basis without invoking remote collapse:
\begin{equation}
\left( \hat{\Pi}_A(\theta_s) \otimes \hat{I}_B \right) [v_{AB}] = |\theta_s\rangle_A \otimes |\chi_B(\theta_s)\rangle,
\label{eq:coset_slice}
\end{equation}
where the conditional coset slice $|\chi_B(\theta_s)\rangle$ of the unmeasured idler field is given by Eq.~(\ref{eq:chi_B_secB}) (see Appendix~\ref{app:measurement} for the dual-bra projection and the three-scan phase equivalence) \cite{Yoon2026b}.

The operational single-photon counting rate $R_D(\theta_s)$ registered at the signal single-photon counting detector SPD ($\approx 807\text{ nm}$) [Fig.~\ref{fig:fig1}] is determined by the squared norm of the projected coset slice \cite{Lee2018,Yoon2021,Glauber1963}:
\begin{align}
R_D(\theta_s) &\propto \big\| \left( \hat{\Pi}_A(\theta_s) \otimes \hat{I}_B \right) [v_{AB}] \big\|^2 = \langle \chi_B(\theta_s) | \chi_B(\theta_s) \rangle \notag \\
&= \frac{1}{2} \left[ |c_1|^2 + |c_2|^2 + c_1 c_2^* \langle m_2 | m_1 \rangle_i e^{i\theta_s} + \text{c.c.} \right] \notag \\
&= \frac{1}{2} \left[ 1 + 2 |c_1 c_2| F \cos\left( \Delta\phi_s - \Delta\phi_p + \Delta\phi_{\text{seed}} \right) \right],
\label{eq:counting_rate_deriv}
\end{align}
whose argument is the scalar invariant $\Delta\Phi$ of Eq.~(\ref{eq:total_phase_secB}). Equation~(\ref{eq:counting_rate_deriv}) demonstrates that the appearance of interference fringes does not require non-local collapse of the idler state $|\chi_B\rangle$ \cite{Yoon2026b}. Instead, the measured rate is the geometric trace over the local coset slice, reconciling relativistic causality with quantum nonlocality \cite{Lee2018,Yoon2021,Yoon2026b}.

\subsection{Closed-Form Quantitative Complementarity and Triality Relations}
\label{subsec:complementarity}

The observable fringe visibility $V$, quanton source purity $\mu_s$, a priori path predictability $P$, and quanton-environment entanglement $E$ are derived in closed algebraic form from Eqs.~(\ref{eq:reduced_density})--(\ref{eq:F_def}) \cite{Yoon2021,Qian2020,Jakob2010}:
\begin{align}
P &= |\rho_{11} - \rho_{22}| = \frac{|\alpha|^2(1 - \gamma^2)}{2 + |\alpha|^2(1+\gamma^2)}, \label{eq:P_closed} \\
V &= 2|\rho_{12}| = V_0 F = \frac{2\gamma|\alpha|^2}{2 + |\alpha|^2(1+\gamma^2)}, \label{eq:V_closed} \\
E &= 2\sqrt{\rho_{11}\rho_{22}(1 - F^2)} = \frac{2\sqrt{1 + |\alpha|^2(1+\gamma^2)}}{2 + |\alpha|^2(1+\gamma^2)}, \label{eq:E_closed}
\end{align}
where $V_0 = 2\sqrt{\rho_{11}\rho_{22}} = \frac{2\sqrt{(1+|\alpha|^2)(1+\gamma^2|\alpha|^2)}}{2+|\alpha|^2(1+\gamma^2)}$ denotes the intrinsic a priori visibility \cite{Greenberger1988,Yoon2021}.

The source purity $\mu_s$, defined through the trace of the squared reduced density matrix $\mu_s \equiv \sqrt{2\operatorname{Tr}[\hat{\rho}_s^2] - 1}$ \cite{Qian2020,Yoon2021}, satisfies the exact Pythagorean relation:
\begin{equation}
P^2 + V^2 = \mu_s^2 = 1 - E^2.
\label{eq:duality_pythagorean}
\end{equation}
Summing the terms yields the unified wave-particle-entanglement triality equality \cite{Yoon2021,Jakob2010}:
\begin{equation}
P^2 + E^2 + V^2 = 1.
\label{eq:triality_equality}
\end{equation}
Equation~(\ref{eq:triality_equality}) shows that the which-way distinguishability $D \equiv \sqrt{P^2 + E^2} = \sqrt{1 - V^2}$ is partitioned between quantum entanglement ($E$) with the undetected idler environment and classical beam imbalance ($P$) \cite{Yoon2021}. When the seed beams are balanced ($\gamma = 1$), the predictability vanishes ($P = 0$), reducing the complementarity relation to $V = F = |\alpha|^2 / (1 + |\alpha|^2)$ and $E = \sqrt{1 - F^2} = \sqrt{1 + 2|\alpha|^2} / (1 + |\alpha|^2)$, which establishes the seed photon number $|\alpha|^2$ as a direct knob controlling environmental entanglement and path indistinguishability [as demonstrated by the measured visibility and decoherence data in Fig.~\ref{fig:fig2}(d); the continuous parameter-space surfaces are shown in Supplemental Material, Fig.~\ref{fig:figS1}] \cite{Lee2018,Yoon2021}.

\subsection{Hardware-Anchored Time-Frequency Stabilizers and Continuous-Variable Grid Embedding}
\label{subsec:gkp_stabilizers}

Because each spectral line $\omega_{s,n}$ in the supermode $\hat{S}_j^\dagger$ is referenced to the atomic-clock-stabilized optical frequency comb [as illustrated in the experimental layout of Fig.~\ref{fig:fig1}] \cite{Lee2018,Udem2002,Diddams2000}, the dual-crystal geometry inherently establishes the displacement lattice of a continuous-variable time-frequency Gottesman-Kitaev-Preskill (TF-GKP) code in the longitudinal chronocyclic phase space \cite{Fabre2020,Fabre2022a,Descamps2024,Yoon2026a}. The periodic comb teeth map the continuous temporal and spectral degrees of freedom onto a discrete, phase-locked lattice, so that the stabilizer and logical translations of the code are fixed by the repetition rate rather than by an auxiliary squeezing or cavity resource. We make this encoding explicit here, because the distinction between a stabilizer cell and a logical cell determines which comb operations are stabilizers, which are logical gates, and how large the correctable displacement region is.

\begin{figure}[t]
    \centering
    \includegraphics[width=\textwidth]{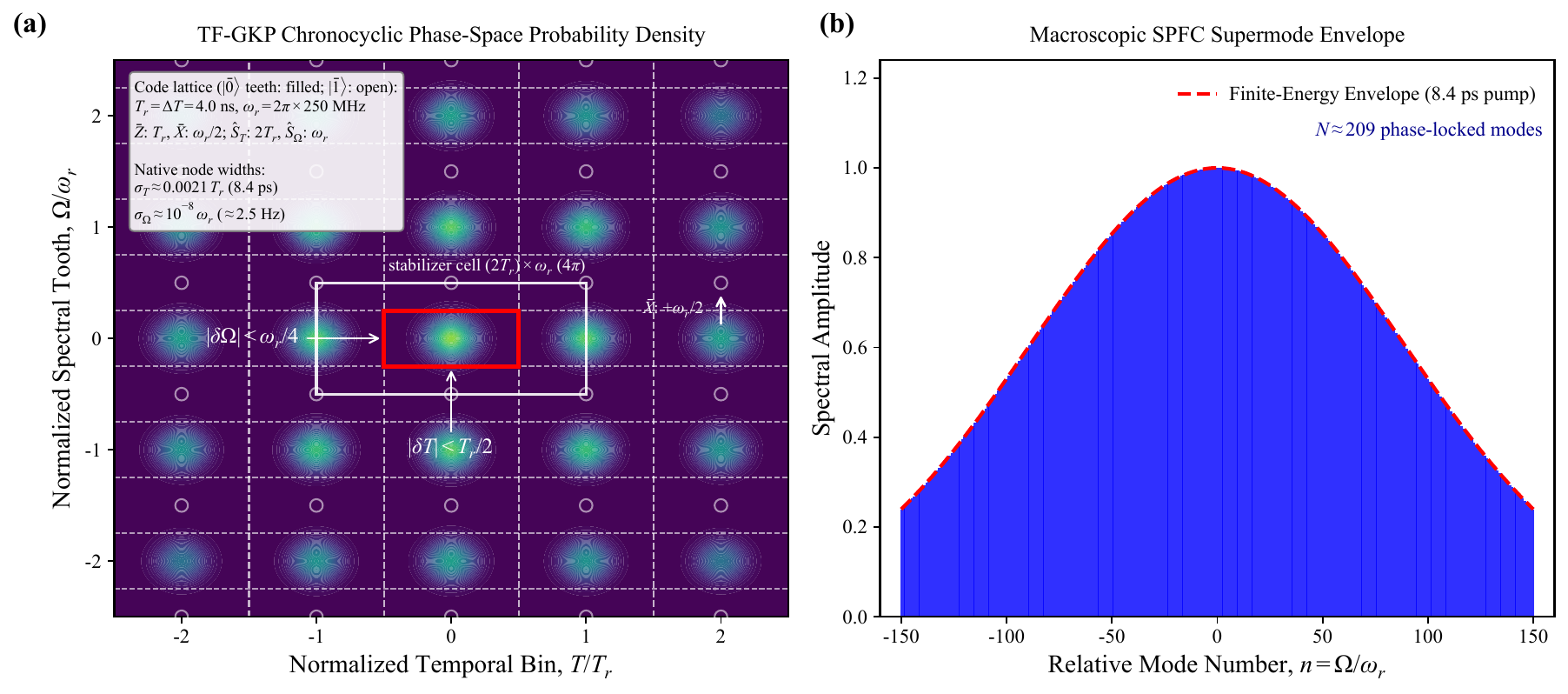}
    \caption{\textbf{Hardware-native continuous-variable Time-Frequency Gottesman-Kitaev-Preskill (TF-GKP) logical qubit framework.} \textbf{(a)} The TF-GKP chronocycle phase space probability density of the SPFC supermode. The physical coordinates are strictly normalized to the temporal bin interval $T_r = \Delta T = 4.0$~ns and the spectral grid spacing $\omega_r = 2\pi \times 250$~MHz. Filled nodes are the teeth of the single-seed codeword $|\bar{0}\rangle$; open circles mark the unpopulated tooth positions of $|\bar{1}\rangle$, displaced by $\omega_r/2$. The dashed grid outlines the lattice of the logical translations $\bar{Z}$ (period $T_r$) and $\bar{X}$ (period $\omega_r/2$), Eq.~(\ref{eq:gkp_logicals}); the white rectangle is one cell of the stabilizers $\hat{\mathcal{S}}_T$ and $\hat{\mathcal{S}}_\Omega$, Eq.~(\ref{eq:gkp_stabilizers}), of periods $2T_r$ and $\omega_r$ and symplectic area $4\pi$. The solid red rectangle highlights the Voronoi cell of the code; environmental shifts bounded by $|\delta T| < T_r/2$ and $|\delta \Omega| < \omega_r/4$ act as tolerated continuous displacements, whereas a shift by $\omega_r/2$ is a logical $\bar{X}$ error. The extreme localization of the physical grid parameters ($\sigma_T \approx 0.0021 T_r$ governed by the $8.4$~ps pump duration, and $\sigma_\Omega \approx 10^{-8} \omega_r$ governed by the sub-Hz cw seed) places the finite-energy codeword deeply inside this cell (plotted probability density nodes are log-scaled for visual clarity). \textbf{(b)} The macroscopic SPFC supermode envelope in the frequency domain. The finite-energy spectral envelope, parameterized by the $8.4$~ps pump, spans $N \approx 209$ discrete phase-locked comb teeth. This illustrates how the comb-driven source intrinsically regularizes the unphysical infinite GKP grid into a realizable, finite-energy codeword $|\bar{0}\rangle$; the codeword $|\bar{1}\rangle$ is the same envelope displaced by $\omega_r/2$.}
    \label{fig:fig4}
\end{figure}

Defining the continuous modal quadratures $\hat{T} = \int t\, |t\rangle\langle t|\, dt$ and $\hat{\Omega} = \int \omega\, |\omega\rangle\langle\omega|\, d\omega$ with $[\hat{T}, \hat{\Omega}] = i$ \cite{Fabre2020,Yoon2026a}, the single-photon frequency comb state $|\Psi_{\text{SPFC}}\rangle = \hat{S}_j^\dagger |0\rangle$ is a finite-energy comb in $\hat{\Omega}$ with teeth at $\Omega_0 + n\omega_r$ and a Fourier-dual pulse train in $\hat{T}$ of period $\Delta T$, whose envelope spans $N \approx 209$ phase-locked modes [Fig.~\ref{fig:fig4}(b)] \cite{Gottesman2001,Yoon2026a}. The two displacements that leave this comb invariant,
\begin{equation}
\hat{\mathcal{Q}}_\tau = \exp\!\left[ -i \Delta T\, (\hat{\Omega}-\Omega_0) \right], \quad \hat{\mathcal{Q}}_\omega = \exp\!\left( i \omega_r \hat{T} \right),
\label{eq:qunaught_lattice}
\end{equation}
commute because $\Delta T\, \omega_r = 2\pi$, but their primitive cell has symplectic area $2\pi$ and therefore supports a one-dimensional common eigenspace: they are the stabilizers of a GKP \emph{qunaught} (sensor) lattice \cite{Walshe2020}, not of a qubit code. A logical qubit requires a stabilizer cell of area $4\pi$. Following the chronocyclic construction of Ref.~\cite{Yoon2026a}, we therefore choose the logical translation scales $T_L = \Delta T$ and $\Omega_L = \omega_r/2$, with $T_L\Omega_L = \pi$, and define the logical Pauli translations
\begin{equation}
\bar{Z} = \exp\!\left[ -i T_L\, (\hat{\Omega}-\Omega_0) \right], \qquad \bar{X} = \exp\!\left( i \Omega_L \hat{T} \right),
\label{eq:gkp_logicals}
\end{equation}
which anticommute, $\bar{X}\bar{Z} = e^{-iT_L\Omega_L}\bar{Z}\bar{X} = -\bar{Z}\bar{X}$, as required. The stabilizers of the TF-GKP code are the doubled translations
\begin{equation}
\begin{split}
\hat{\mathcal{S}}_T &= \bar{Z}^2 = \exp\!\left[ -2i \Delta T\, (\hat{\Omega}-\Omega_0) \right], \\
\hat{\mathcal{S}}_\Omega &= \bar{X}^2 = \exp\!\left( i \omega_r \hat{T} \right),
\end{split}
\label{eq:gkp_stabilizers}
\end{equation}
with $[\hat{\mathcal{S}}_T, \hat{\mathcal{S}}_\Omega] = 0$ and cell area $(2\Delta T)\,\omega_r = 4\pi$. Comparing Eqs.~(\ref{eq:qunaught_lattice}) and (\ref{eq:gkp_stabilizers}), $\hat{\mathcal{Q}}_\omega = \hat{\mathcal{S}}_\Omega$ is a genuine stabilizer, whereas $\hat{\mathcal{Q}}_\tau = \bar{Z}$ is a logical operator; this corrects the identification of both operators as stabilizers in the preprint version of this work (arXiv:2610.27602v1) and makes the present construction consistent with the stabilizer geometry of Ref.~\cite{Yoon2026a}.

The code space is the common $+1$ eigenspace of Eq.~(\ref{eq:gkp_stabilizers}). The condition $\hat{\mathcal{S}}_T|\bar{\psi}\rangle = |\bar{\psi}\rangle$ restricts the spectral support to the interleaved grid $\Omega = \Omega_0 + \mu\,\omega_r/2$, $\mu \in \mathbb{Z}$, and the two cosets of this grid are the ideal codewords
\begin{equation}
|\bar{0}\rangle \propto \sum_{n} \big|\Omega_0 + n\omega_r\big\rangle, \qquad |\bar{1}\rangle \propto \sum_{n} \big|\Omega_0 + (n+\tfrac{1}{2})\omega_r\big\rangle,
\label{eq:gkp_codewords}
\end{equation}
so that $\bar{Z}|\bar{b}\rangle = (-1)^b|\bar{b}\rangle$ and $\bar{X}|\bar{0}\rangle = |\bar{1}\rangle$. The SPFC supermode generated with a single seed frequency, Eq.~(\ref{eq:supermode_def}), is a finite-energy realization of the codeword $|\bar{0}\rangle$: its teeth sit on the integer grid, and its temporal form is a pulse train of period $\Delta T$ in which the carrier phase advances by $2\pi\Omega_0/\omega_r$ from pulse to pulse. The codeword $|\bar{1}\rangle$ is the same comb displaced by half a tooth, i.e.\ the pulse train with an additional carrier-phase slip of $\pi$ per round trip; it is produced by shifting the seed frequency (or, equivalently, the pump carrier-envelope offset) by $f_r/2$ with an acousto-optic or electro-optic single-sideband modulator. In the time domain, the $\bar{X}$ eigenstates $|\bar{\pm}\rangle$ are pulse trains occupying only the even or only the odd time slots of length $\Delta T$. The logical operations are therefore native comb operations: $\bar{Z}$ is a delay by one repetition period, $\bar{X}$ is a frequency shift by half a tooth spacing (a serrodyne drive at $f_r/2$), and the stabilizers are a delay by $2\Delta T$ and a shift by one full tooth. More generally, any spectral phase filter periodic in $\Omega$ with period $\omega_r$, and any temporal phase modulation periodic in $t$ with period $2\Delta T$, commutes with both stabilizers and acts inside the code space as a rotation about $\bar{Z}$ or $\bar{X}$, respectively, so that the complete single-qubit gate set is available as passive or electro-optic linear optics.

A bichromatic seed with components $\alpha_1$ at $\omega_{\text{seed}}$ and $\alpha_2$ at $\omega_{\text{seed}} + \omega_r/2$ in a \emph{single} crystal populates both cosets at once. Because $\omega_{s,n} = \omega_{p,n} - \omega_i$, the two seed components generate the two codewords, and the conditional single-photon state takes exactly the form of Eq.~(\ref{eq:psi_ENBS}) with the path labels $|1,0\rangle_s$ and $|0,1\rangle_s$ replaced by $|\bar{0}\rangle$ and $|\bar{1}\rangle$,
\begin{equation}
|\Psi_1\rangle_{L} = c_1\, |\bar{0}\rangle\, |m_1\rangle_i + c_2\, |\bar{1}\rangle\, |m_2\rangle_i,
\label{eq:logical_state_bichromatic}
\end{equation}
with the same normalization coefficients, Eq.~(\ref{eq:norm_coeff}), and the same single-photon-added coherent idler states, Eq.~(\ref{eq:m_states}), now attached to the two seed components rather than to the two crystals. The logical amplitudes are set by the seed balance $\gamma$ and the logical phase by $\Delta\phi_{\text{seed}} - \Delta\phi_p$, so that one crystal with one frequency-shifted seed component addresses the full logical Bloch sphere. The equal-weight state with zero relative phase is $|\bar{+}\rangle$, a comb on the half-integer grid, i.e.\ the qunaught state of the dual rectangular lattice $(2\Delta T, \omega_r/2)$, whereas the single-seed codeword $|\bar{0}\rangle$ is the qunaught state stabilized by Eq.~(\ref{eq:qunaught_lattice}); a relative phase of $\pi/4$ gives the magic state $(|\bar{0}\rangle + e^{i\pi/4}|\bar{1}\rangle)/\sqrt{2}$ at no additional cost, because in this single-photon modal encoding a non-Clifford state differs from a Clifford one only by a seed phase. These statements concern the single-photon component of the seeded output, conditioned on the registration of one signal photon. The unconditioned stimulated field is a displaced thermal state whose coherent part dominates for $|\alpha|^2 \gg 1$; its photon statistics are Poissonian, so a one-photon time bin is contaminated by two photons with probability $P_2/P_1 = \bar{n}/2$, and the idler record carries only the information $E^2 = 1 - F^2$ ($\approx 0.02$ at $|\alpha|^2 = 100$) about whether a signal photon was emitted, so the seeded carrier cannot be heralded. The seeded source is therefore a bright, phase-coherent twin of the qubit mode, with $\bar{n}$ tunable from $10^{-2}$ to $10^{2}$ photons per pulse, suited to alignment, calibration and logical tomography of the encoding; deterministic single-photon preparation of the same states requires the unseeded signal--idler pair with a herald on the partner photon, which lies outside the present experiments. Tracing out the idler as in the section \emph{Open Quantum System Formulation: Tracing Out the Undetected Continuum} shows that the purity of the encoded qubit is governed by the same SPACS overlap $F$ of Eq.~(\ref{eq:F_def}): the triality relation of Eq.~(\ref{eq:triality_equality}) is, in this reading, the statement that logical predictability (seed imbalance), logical coherence $V = V_0 F$, and entanglement with the unobserved idler exhaust the Bloch vector of the encoded qubit, and the spatial dephasing rate $\Gamma_{\text{deph}} = -(1/L_c)\ln F$ is the rate at which the logical coherence is lost to the idler environment inside the crystal.

In the dual-crystal geometry of Fig.~\ref{fig:fig1}, the same half-tooth shift applied to the pump or the seed of one crystal only correlates the which-crystal path qubit with the TF-GKP qubit, $|\Psi_1\rangle = c_1|1,0\rangle_s|\bar{0}\rangle|m_1\rangle_i + c_2|0,1\rangle_s|\bar{1}\rangle|m_2\rangle_i$. The local coset projection $\hat{\Pi}_A(\theta_s)$ of Eq.~(\ref{eq:PVM_local}) then no longer produces an intensity fringe, because $|\bar{0}\rangle$ and $|\bar{1}\rangle$ are orthogonal; instead, it transfers the path coherence into the logical qubit and, in the macroscopic-seed limit $F \to 1$, leaves on the detected output port the pure state $c_1|\bar{0}\rangle + c_2\, e^{i(\theta_s - \Delta\phi_p + \Delta\phi_{\text{seed}})}|\bar{1}\rangle$, the other port carrying the same state with $\theta_s \to \theta_s + \pi$; for $F < 1$ the logical coherence is reduced by the factor $F$, exactly as in Eq.~(\ref{eq:V_closed}). The three-scan phase equivalence of Appendix~\ref{app:measurement} thus becomes a gate identity: the logical phase can be applied through the pump, the seed, or the detector, and the quotient-space statement that the correlations are pre-encoded in $[v_{AB}]$ is the statement that this encoding step requires no remote action on the idler.

Small environmental timing jitter and frequency offsets manifest as continuous displacements and are tolerated provided they remain inside the Voronoi cell of the code, $|\delta t| < \Delta T/2 = 2.0$~ns and $|\delta\omega| < \omega_r/4 = 2\pi \times 62.5$~MHz [Fig.~\ref{fig:fig4}(a)]; a frequency shift by $\omega_r/2$ is a logical $\bar{X}$ error rather than a correctable displacement, which halves the spectral bound relative to the qunaught cell of Eq.~(\ref{eq:qunaught_lattice}). With the $8.4$-ps pump localization ($\sigma_T \approx 2\times 10^{-3}\,\Delta T$) and Hertz-level tooth linewidths ($\sigma_\Omega \approx 10^{-8}\,\omega_r$), both bounds are met by many orders of magnitude, and the finite-energy overlap $\langle\bar{0}|\bar{1}\rangle \sim \exp[-(\omega_r/4\sigma_\Omega)^2]$ is negligible. We stress the scope of this protection. The TF-GKP structure absorbs unitary displacement errors (timing jitter, frequency drift, dispersion) passively at readout, a property not shared by path, polarization, or time-bin qubits, and it renders every single-qubit operation a deterministic linear-optical element; it does not correct photon loss, and it does not by itself supply an entangling gate between photons. Those must be provided by an outer loss-tolerant code and by measurement-induced (fusion) entanglement \cite{Bartolucci2023}, as discussed in the section \emph{Outlook and Direct Impacts on Quantum Information Science and Technology}.

\section{Discussion}
\label{sec:discussion_conclusion}

\subsection{Physical Benchmarks of the Spacelike-Separated Dual PPLN Architecture}
\label{subsec:unique_capabilities}

The theoretical framework established in the sections \emph{Measurement of Quantum States and Nonlocal Correlations} and \emph{Theoretical Framework} accounts for several physical benchmarks realized in our dual-crystal single-photon frequency-comb experiments [see experimental layout in Fig.~\ref{fig:fig1}] that cannot be duplicated by standard single-crystal parametric down-conversion or passive linear optics \cite{Lee2018,Yoon2021}.

First, rather than generating single photons at a single location and probabilistically splitting them across two spatial paths with a passive beam splitter, our system generates signal single-photon frequency-comb supermodes from two physically distinct dielectric media separated across a laboratory baseline $L_0 \approx 1\text{ m}$ \cite{Lee2018,Yoon2021}. The local interaction volumes satisfy relativistic microcausality $[\hat{\mathbf{A}}_\perp(x_1), \hat{\mathbf{A}}_\perp(x_2)] = 0$ [Eq.~(\ref{eq:microcausality})] \cite{Lee2018,Peskin1995}. We observe unheralded single-particle interference fringes with visibility up to $V \approx 0.99$ without detecting the conjugate idler modes, confirming that \textit{coherence is induced purely through the quantum indistinguishability of the two non-overlapping sources} rather than classical wave division \cite{Lee2018,Qian2020,Yoon2021,Yoon2026b}.

Second, our experiments directly verify the \textit{operational three-scan phase equivalence} [Fig.~\ref{fig:fig2}(a) and \ref{fig:fig2}(b), and Appendix~\ref{app:measurement}] \cite{Lee2018,Yoon2026b}. Independently scanning the pump comb path difference ($\Delta x_p$), the cw idler seed path difference ($\Delta x_{\text{seed}}$), or the signal interferometric path difference ($\Delta x_s$) produces identical sinusoidal fringes with near-unit contrast, each with spatial periodicity corresponding to its respective wavelength ($\lambda_p = 530\text{ nm}$, $\lambda_i = 1542.38\text{ nm}$, and $\lambda_s = 807.2\text{ nm}$) \cite{Lee2018,Yoon2026b}. This equivalence confirms that single-particle interference is governed by the relative phase between the composite quantum state and the detector-selected measurement basis \cite{Lee2018,Yoon2026b}.

Third, by adjusting the seed beam amplitude ratio $\gamma = |\alpha_1|/|\alpha_2|$ and seed photon number $|\alpha|^2$, our system allows \textit{continuous experimental control over the SPACS state overlap fidelity} $F \in [0, 1]$ [Eq.~(\ref{eq:F_def})] \cite{Lee2018,Yoon2021}. This capability has enabled the complete experimental mapping and verification of the wave-particle-entanglement triality relation $P^2 + E^2 + V^2 = 1$ and the source purity bound $\mu_s = \sqrt{1 - E^2}$ across their entire two-dimensional parameter space [Fig.~\ref{fig:fig2}(c) and \ref{fig:fig2}(d)], demonstrating that which-path distinguishability is fundamentally partitioned between classical beam imbalance and quantum entanglement with the unobserved idler reservoir \cite{Jakob2010,Yoon2021,Qian2020}.

\subsection{Foundational Status and Experimental Open Questions}
\label{subsec:remaining_issues}

While our open-system QED formulation resolves how definite measurement statistics emerge from local coset projections without invoking non-local superluminal collapse, several conceptual questions remain open when compared against other modern foundational frameworks \cite{Wiseman2010,Scholes2026,Brunner2014}.

The free-vector-space quotient formulation introduced by Scholes \cite{Scholes2026,Scholes2026a} and physically realized here reveals that an entangled state vector intrinsically encodes an ensemble of measurement outcomes via its equivalence class representatives mapped from the overarching free vector space $F(V_A \times V_B)$ \cite{Scholes2026}. This model explains why joint measurement outcomes between the two spatially separated crystal volumes ($\mathrm{NL}_1$ and $\mathrm{NL}_2$) are correlated and why local subsystem vectors simplify via destructive and constructive interference into definite eigenstates upon projection \cite{Scholes2026,Scholes2026a}. However, it remains an open question at what stage an individual experimental run actualizes one specific eigenvalue out of the statistical ensemble without a dynamical reduction mechanism (such as objective gravitational collapse models \cite{Penrose1998} or many-worlds branchings) \cite{Wiseman2010,Zurek2003}. As highlighted by modern analyses of extended Wigner's friend experiments regarding the non-absoluteness of observed events \cite{Zwirn2025}, reconciling universal unitary dynamics with definite measurement outcomes requires careful consideration of perspectival frameworks. While our open-system Lindblad formulation maps ensemble decoherence via the idler continuum, the single-shot actualization and the emergence of absolute events remain outside standard unitary dynamics \cite{Lindblad1976,Zurek2003}.

Furthermore, in our 1-m baseline setup, the optical time of flight ($\sim 3.3\text{ ns}$) is comparable to the 4-ns pulse repetition interval \cite{Lee2018}. Although spatial microcausality holds at the operator level, strict verification of Bell nonlocality without the locality and freedom-of-choice loopholes requires integrating high-speed electro-optic phase modulators driven by quantum random number generators operating outside each other's forward light cones \cite{Aspect1988,Brunner2014}. Distributing the dual-source comb architecture across kilometer-scale fiber networks or free-space optical links will be essential to test contextual quotient projection under strict Einstein locality \cite{Lee2018,Brunner2014}---a macroscopic network extension that lies beyond the scope of the present study and is reserved for future experimental investigation.

Finally, as described in Sec.~\ref{sec:methods}, our current single-photon detection apparatus employs an EMCCD-coupled spectrometer operating with a 10-ms integration window, measuring ensemble interference fringes averaged over $2.5 \times 10^6$ pulses at the deep single-photon filling factor of $\sim 0.02$ photons per pulse \cite{Lee2018}. Although the Hertz-level linewidth of individual teeth is verified in the high-flux regime via RF heterodyne beat-note spectroscopy, resolving single-photon comb teeth in real time remains a technological barrier \cite{Lee2018}. Complementary platforms utilizing electro-optic quantum pulse gates or nonlinear sum-frequency dispersion sorters offer a viable path to perform full comb-tooth-resolved tomography on individual SPFC supermodes \cite{Brecht2015,Lu2018,Serino2023}. For the encoded qubit of the section \emph{Hardware-Anchored Time-Frequency Stabilizers and Continuous-Variable Grid Embedding} the required readout is simpler than full tomography: $\bar{Z}$ is read by sorting integer from half-integer teeth with a resonator of free spectral range $f_r$, $\bar{X}$ by time-tagging the photon modulo $2\Delta T$, and rotated bases by a quantum pulse gate whose shaped pump is derived from the same comb; balanced homodyne detection is not a substitute, because the quadrature statistics of a one-photon state are independent of the local-oscillator phase.

\subsection{Outlook and Direct Impacts on Quantum Information Science and Technology}
\label{subsec:outlook_impacts}

The synthesis of dual-crystal stimulated parametric down-conversion [Fig.~\ref{fig:fig1}], optical frequency comb metrology [Eqs.~(\ref{eq:pump_comb})--(\ref{eq:supermode_def})], and open-system measurement theory [the microcausality condition of Eq.~(\ref{eq:microcausality}) and the dephasing framework of Eq.~(\ref{eq:lindblad})] yields practical dividends across several pillars of quantum technology.

In photonic quantum computing with hardware-native TF-GKP qubits, anchoring the single-photon frequency comb to an atomic-clock-referenced OFC fixes the stabilizer lattice $(2\Delta T, \omega_r)$ and the logical translations $(\Delta T, \omega_r/2)$ of the code in the longitudinal time-frequency phase space [Eqs.~(\ref{eq:gkp_logicals})--(\ref{eq:gkp_stabilizers})] \cite{Gottesman2001,Yoon2026a}. Because the discrete supermode operator $\hat{S}_j^\dagger$ of Eq.~(\ref{eq:supermode_def}) spans $N \approx 209$ phase-locked modes with Hertz-level individual linewidths, the single-photon component of the seeded output is the finite-energy codeword $|\bar{0}\rangle$ of this code, and a seed component shifted by $f_r/2$ steers it to $|\bar{1}\rangle$, $|\bar{\pm}\rangle$ and the magic state with the same source \cite{Lee2018,Yoon2026a}. Logical Pauli operations are a one-period delay and a half-tooth frequency shift, and arbitrary single-qubit rotations are comb-periodic spectral or temporal phases, all realizable with passive or electro-optic linear optics; the protection the code provides is passive tolerance of timing, frequency and dispersion errors inside the Voronoi cell. Entangling operations, by contrast, are not provided by the encoding, and neither are heralded single photons by the seeded source: both must come from the signal--idler pair generated by the same crystal in the unseeded regime, which realizes the collective two-photon TF-GKP code of Ref.~\cite{Descamps2024} (a comb in the sum frequency) and, when the crystal is placed in a comb-locked cavity of free spectral range $f_r/2$ that imposes the interleaved grid on each photon, becomes a Bell pair of single-mode TF-GKP qubits whose parity is set by the pump carrier-envelope offset; measurement-induced fusion between such photons \cite{Bartolucci2023} then supplies the remaining entanglement, with photon loss handled by the outer fusion network rather than by the TF-GKP layer. In this sense the architecture provides the displacement-tolerant encoding, with the seeded source as its calibration twin and the unseeded, heralded pair as its qubit source, for measurement-based and fusion-based photonic quantum computing \cite{Bourassa2021,Bartolucci2023,Larsen2025}, building upon foundational demonstrations of macroscopic continuous-variable entanglement in frequency \cite{Yin2014} and time \cite{Yokoyama2013} domains.

In high-dimensional quantum key distribution (QKD) and quantum communication, the high-dimensional Hilbert space spanned by the $N \approx 209$ comb modes offers an ideal substrate for energy-time and frequency-bin qudit protocols \cite{Xie2015,Kues2017,Lu2023a}. Encoding cryptographic keys within the discrete supermode comb teeth enables high secret-key capacities per detected photon while maintaining high resilience against channel decoherence during transmission through standard telecommunication fibers \cite{Lee2018,Yoon2026a}. Furthermore, because the single-photon coherence of the signal field is controlled remotely via the undetected idler seed parameters ($\gamma, |\alpha|^2$), this dual-source setup enables unheralded, remote quantum state engineering and quantum routing across quantum networks \cite{Lee2018,Yoon2021}.

In quantum metrology and quantum sensing with undetected photons, the sensitivity of the SPFC fringe visibility to the idler state overlap ($V = V_0 F$) establishes a platform for quantum spectroscopy and quantum imaging with undetected photons \cite{Lemos2014,Lee2017,Lee2018,Lee2019,Lee2020}.
Placing an uncharacterized optical medium or molecular sample into the 1542-nm idler beam path modulates the transmission $T(\omega_i)$ and phase shift $\Delta\phi_{\text{seed}}$, which is directly read out through the visibility and position of the 807-nm single-photon signal interference fringes on an ordinary silicon EMCCD camera \cite{Lee2018,Lee2019,Lee2020}. This capability eliminates the need for noisy or inefficient mid-infrared photodetectors, enabling high-precision infrared spectroscopy, trace chemical sensing, and quantum optical coherence tomography with visible-wavelength detection \cite{Lee2018,Lee2019,Lee2020,Descamps2023a}.

In summary, the relativistic QED open-system framework and quotient-space measurement theory presented here unify the physics of induced coherence, Bohr's complementarity, and continuous-variable bosonic grid stabilization \cite{Lee2018,Yoon2021,Yoon2026a,Scholes2026}. By confirming that single-photon quantum interference is defined jointly by the source quantum state and the detector measurement slice, this work bridges foundational tests of quantum nonlocality and the realization of displacement-tolerant TF-GKP logical qubits for photonic quantum information processing \cite{Lee2018,Yoon2021,Yoon2026a,Yoon2026b}.

\section{Materials and Methods}
\label{sec:methods}
\paragraph*{Dual-crystal frequency-comb biphoton source.} The two nonlinear biphoton sources are identical periodically poled lithium niobate (PPLN) crystals ($15\times7.9\times0.5$~mm$^3$, grating period 7.3~$\mu$m, type-0 phase matching at $121.5~^\circ$C) separated by $L_0\approx1$~m. Both are pumped by a 530-nm optical frequency comb obtained by frequency doubling a 1060-nm comb (46-nm bandwidth) in LBO; the pump comb ($\sim$3~nm width, $\sim$14~mW) is phase-locked at $f_r=250.0$~MHz and $f_{\rm ceo}=20.0$~MHz to a GPS-disciplined Rb clock \cite{Lee2018}. The cw seed at 1542.384~nm (linewidth $<10$~Hz) is injected along the pump paths with dichroic mirrors. Pump and seed are focused with $f=100$~mm lenses to $w_0\approx17.3~\mu$m, giving a confocal parameter $b\approx7.9$~mm that sets the effective interaction length $L_c$. The signal single-photon supermodes at 807.2~nm are recombined on a 50:50 beam splitter and recorded with an EMCCD-coupled spectrometer (0.1-nm resolution, 10-ms integration), averaging $2.5\times10^6$ pulses at $\approx0.02$ photons per pulse. Pump, seed, and signal phases are scanned independently with piezo-mounted mirrors. Further details are given in Supplemental Material, Sec.~\ref{sec:S1}.

\paragraph*{Theory and numerics.} The spatial Lindblad master equation, the quotient-space measurement construction, and the three-scan phase equivalence are derived in Appendices~\ref{app:lindblad}--\ref{app:qed}. The triality surfaces were evaluated with Mathematica and Python (Supplemental Material, Sec.~\ref{sec:S2}).

\paragraph*{Data, materials, and software availability.} All data needed to evaluate the conclusions are present in the paper, its appendices, and the Supplemental Material. Experimental data in Fig.~\ref{fig:fig2} are adapted from Ref.~\cite{Lee2018}.

\paragraph*{Competing interests.} The author declares no competing interest.

\begin{acknowledgments}
The author thanks S. K. Lee and M. Cho for long-term collaboration on single-photon frequency-comb generation, entangled nonlinear biphoton sources, the quantitative complementarity of wave--particle duality, and phase-sensitive single-photon interferometry. This work was supported by the National Research Foundation of Korea (NRF) under Grant No.~RS-2022-NR068815 and in part by IBS-R023-D1.
\end{acknowledgments}

\appendix
\section{Microscopic Derivation of the Spatial Lindblad Master Equation and Dephasing Rate}
\label{app:lindblad}
\label{app:lindblad_derivation}

In this appendix, we provide the step-by-step derivation of the spatial Lindblad master equation [Eq.~(\ref{eq:lindblad})] and the effective spatial dephasing rate $\Gamma_{\text{deph}}$ [Eq.~(\ref{eq:dephasing_rate})] starting from the discrete Kraus representation.

In the single-photon path subspace $\mathcal{H}_s = \operatorname{span}\{|1,0\rangle_s, |0,1\rangle_s\}$, the non-unitary map arising from the undetected idler modes across an incremental crystal propagation segment $\delta y$ is described by the Kraus operators $\hat{M}_0$ and $\hat{M}_1$ defined in Eq.~(\ref{eq:kraus_ops}). Using the Pauli operator $\hat{\sigma}_z = |1,0\rangle_s\langle 1,0| - |0,1\rangle_s\langle 0,1|$ and the identity $\hat{I} = |1,0\rangle_s\langle 1,0| + |0,1\rangle_s\langle 0,1|$, the identity-like operator $\hat{M}_0$ and the dephasing jump operator $\hat{M}_1$ can be expressed algebraically as:
\begin{align}
\hat{M}_0(\delta y) &= \frac{1+F(\delta y)}{2}\hat{I} + \frac{1-F(\delta y)}{2}\hat{\sigma}_z, \label{eq:app_M0} \\
\hat{M}_1(\delta y) &= \sqrt{1 - F^2(\delta y)}\, \frac{\hat{I} - \hat{\sigma}_z}{2}, \label{eq:app_M1}
\end{align}
where $F(\delta y)$ is the incremental idler overlap fidelity over propagation distance $\delta y$. 

For an incremental propagation distance $\delta y \ll L_c$, we expand the fidelity to first order as $F(\delta y) \approx 1 - \frac{1}{2}\Gamma_{\text{deph}}\delta y + \mathcal{O}(\delta y^2)$. Substituting this expansion into Eqs.~(\ref{eq:app_M0}) and (\ref{eq:app_M1}), the Kraus operators become:
\begin{align}
\hat{M}_0(\delta y) &= \hat{I} - \frac{1}{4}\Gamma_{\text{deph}}\delta y (\hat{I} - \hat{\sigma}_z) + \mathcal{O}(\delta y^2), \\
\hat{M}_1(\delta y) &= \sqrt{\Gamma_{\text{deph}}\delta y}\, \frac{\hat{I} - \hat{\sigma}_z}{2} + \mathcal{O}(\delta y^{3/2}).
\end{align}
The change in the reduced density matrix over $\delta y$ due to dissipation is $\delta\hat{\rho}_s = \hat{M}_0 \hat{\rho}_s \hat{M}_0^\dagger + \hat{M}_1 \hat{\rho}_s \hat{M}_1^\dagger - \hat{\rho}_s$. Evaluating up to order $\delta y$, we obtain:
\begin{equation}
\frac{\delta\hat{\rho}_s}{\delta y} = \frac{\Gamma_{\text{deph}}}{2} \left( \hat{\sigma}_z \hat{\rho}_s \hat{\sigma}_z - \hat{\rho}_s \right) = \mathcal{D}[\hat{\rho}_s],
\label{eq:app_dissipator}
\end{equation}
which is the canonical Lindblad dissipator with jump operator $\hat{L} = \sqrt{\Gamma_{\text{deph}}}\,\hat{\sigma}_z$. Adding the unitary spatial generator $\hat{H}_{\text{prop}} = \hbar k_s \hat{I}$, we arrive directly at the spatial master equation along the longitudinal coordinate $y$:
\begin{equation}
\frac{d\hat{\rho}_s(y)}{dy} = -\frac{i}{\hbar}\left[ \hat{H}_{\text{prop}}, \hat{\rho}_s(y) \right] + \frac{\Gamma_{\text{deph}}}{2}\left( \hat{\sigma}_z \hat{\rho}_s(y)\hat{\sigma}_z - \hat{\rho}_s(y) \right).
\label{eq:app_master_eq}
\end{equation}

In matrix elements, Eq.~(\ref{eq:app_master_eq}) yields $d\rho_{jj}(y)/dy = 0$ ($j \in \{1,2\}$) and $d\rho_{12}(y)/dy = -\Gamma_{\text{deph}}\rho_{12}(y)$. Integrating the off-diagonal coherence across the full physical interaction length $L_c \approx 7.9\text{ mm}$ of the PPLN crystal yields:
\begin{equation}
\rho_{12}(L_c) = \rho_{12}(0) \exp\left(-\Gamma_{\text{deph}} L_c\right).
\label{eq:app_coherence_decay}
\end{equation}
Equating this dynamical solution to the unheralded reduced density matrix [Eq.~(\ref{eq:reduced_density})], where the off-diagonal coherence is scaled by the total idler fidelity $F \equiv |\langle m_2 | m_1 \rangle_i|$ [Eq.~(\ref{eq:F_def})], we identify $\exp\left(-\Gamma_{\text{deph}} L_c\right) = F$. Taking the natural logarithm directly establishes:
\begin{equation}
\Gamma_{\text{deph}} = -\frac{1}{L_c} \ln F.
\label{eq:app_gamma_rate}
\end{equation}

\subsection{Relation between Spatial Dephasing and Temporal Lindblad Rate}
\label{app:spatial_to_temporal}

In conventional open quantum systems, the Lindblad master equation is formulated with respect to time $t$, with transition and decoherence rates expressed in units of frequency ($\mathrm{s}^{-1}$). In contrast, for propagating quantum optical fields through dielectric media, the physical evolution coordinate is the longitudinal propagation distance $y$, yielding a spatial master equation $\frac{d\hat{\rho}_s(y)}{dy}$ whose dephasing rate $\Gamma_{\text{deph}}$ carries the dimension of inverse length ($\mathrm{m}^{-1}$).

The mapping between the spatial propagation coordinate $y$ and laboratory interaction time $t$ is governed by the group velocity of the propagating signal wavepacket inside the PPLN crystal:
\begin{equation}
v_g = \left( \left. \frac{dk_s(\omega)}{d\omega} \right|_{\omega_{s0}} \right)^{-1} = \frac{c}{n_g},
\label{eq:group_velocity}
\end{equation}
where $n_g \approx 2.23$ is the extraordinary group index of lithium niobate at $\lambda_s \approx 807.2\text{ nm}$. Under the paraxial propagation relation $y = v_g t$, the differential operators are related by:
\begin{equation}
\frac{d}{dy} = \frac{1}{v_g} \frac{d}{dt}.
\label{eq:spatial_temporal_diff}
\end{equation}

Multiplying the spatial Lindblad equation [Eq.~(\ref{eq:app_master_eq})] by $v_g$ transforms it into the canonical time-domain Lindblad master equation:
\begin{equation}
\frac{d\hat{\rho}_s(t)}{dt} = -\frac{i}{\hbar} \left[ \hat{H}_{\text{eff}}, \hat{\rho}_s(t) \right] + \frac{\gamma_{\text{deph}}}{2} \left( \hat{\sigma}_z \hat{\rho}_s(t) \hat{\sigma}_z - \hat{\rho}_s(t) \right),
\label{eq:temporal_lindblad}
\end{equation}
where $\hat{H}_{\text{eff}} = v_g \hat{H}_{\text{prop}}$, and the standard temporal dephasing rate $\gamma_{\text{deph}}$ (in $\mathrm{s}^{-1}$ or $\mathrm{rad/s}$) is given by:
\begin{equation}
\gamma_{\text{deph}} = v_g \Gamma_{\text{deph}} = -\frac{v_g}{L_c} \ln F = -\frac{1}{\tau_{\text{int}}} \ln F.
\label{eq:temporal_gamma_rate}
\end{equation}
Here, $\tau_{\text{int}} \equiv L_c / v_g \approx 58.7\text{ ps}$ represents the single-pass transit interaction time of the photon wavepacket traversing the crystal of physical length $L_c = 7.9\text{ mm}$. Equation~(\ref{eq:temporal_gamma_rate}) explicitly demonstrates that the spatial dephasing coefficient $\Gamma_{\text{deph}}$ ($\mathrm{m}^{-1}$) and the temporal rate $\gamma_{\text{deph}}$ ($\mathrm{s}^{-1}$) describe the identical physical decoherence process under the canonical velocity mapping $v_g$.

\section{Free-Vector-Space Quotient Construction and Derivation of the Coset Slice}
\label{app:quotient}
\label{app:quotient_derivation}

In this appendix, we present the algebraic foundations of the free-vector-space quotient construction and derive the explicit form of the conditional coset state $|\chi_B(\theta_s)\rangle$ [Eq.~(\ref{eq:chi_B_secB})].

Let $V_A = \mathcal{H}_s$ and $V_B = \mathcal{H}_i$ denote the Hilbert spaces of the signal and idler fields, respectively. In abstract multilinear algebra, the free vector space $F(V_A \times V_B)$ over the complex field $\mathbb{C}$ is the infinite-dimensional vector space whose basis consists of all Cartesian pairs $(\ket{\phi}_A, \ket{\psi}_B)$ with $\ket{\phi}_A \in V_A$ and $\ket{\psi}_B \in V_B$. An arbitrary element $v \in F(V_A \times V_B)$ is an explicit finite formal linear combination $v = \sum_{k} z_k ( \ket{\phi_k}_A, \ket{\psi_k}_B )$, $z_k \in \mathbb{C}$. In $F(V_A \times V_B)$, elements are formal strings without multilinear relations; for instance, $(c\ket{\phi}_A, \ket{\psi}_B) \neq c(\ket{\phi}_A, \ket{\psi}_B)$ for $c \neq 1$.

To impose bilinearity, one constructs the linear subspace $R \subset F(V_A \times V_B)$ generated by all vectors of the form:
\begin{align}
&(\ket{\phi_1}_A + \ket{\phi_2}_A, \ket{\psi}_B) - (\ket{\phi_1}_A, \ket{\psi}_B) - (\ket{\phi_2}_A, \ket{\psi}_B), \notag \\
&(\ket{\phi}_A, \ket{\psi_1}_B + \ket{\psi_2}_B) - (\ket{\phi}_A, \ket{\psi_1}_B) - (\ket{\phi}_A, \ket{\psi_2}_B), \notag \\
&c(\ket{\phi}_A, \ket{\psi}_B) - (c\ket{\phi}_A, \ket{\psi}_B), \notag \\
&c(\ket{\phi}_A, \ket{\psi}_B) - (\ket{\phi}_A, c\ket{\psi}_B),
\label{eq:app_relations}
\end{align}
for all $\ket{\phi}_A, \ket{\phi_j}_A \in V_A$, $\ket{\psi}_B, \ket{\psi_j}_B \in V_B$, and $c \in \mathbb{C}$. The tensor product Hilbert space $\mathcal{H}_{AB} \cong V_A \otimes V_B$ is defined rigorously as the quotient vector space $\mathcal{H}_{AB} = F(V_A \times V_B) / R$. The canonical quotient map $\pi: F(V_A \times V_B) \to \mathcal{H}_{AB}$ maps each formal basis pair to its equivalence class (coset): $\pi( (\ket{\phi}_A, \ket{\psi}_B) ) \equiv [(\ket{\phi}_A, \ket{\psi}_B)] = (\ket{\phi}_A, \ket{\psi}_B) + R \equiv \ket{\phi}_A \otimes \ket{\psi}_B$.

Before measurement, the composite state generated by the dual PPLN crystals is represented in $F(V_A \times V_B)$ by the formal sum:
\begin{equation}
v_{AB} = c_1 \left( |1,0\rangle_s, |m_1\rangle_i \right) + c_2 \left( |0,1\rangle_s, |m_2\rangle_i \right).
\label{eq:app_state_free}
\end{equation}
Under the quotient map $\pi$, $v_{AB}$ defines the equivalence class $[v_{AB}] = c_1 |1,0\rangle_s \otimes |m_1\rangle_i + c_2 |0,1\rangle_s \otimes |m_2\rangle_i \in \mathcal{H}_{AB}$.

A local projection-valued measurement at the signal detector is defined by $\hat{\Pi}_A(\theta_s) = |\theta_s\rangle_A \langle \theta_s|_A$, with the detector-defined basis state $|\theta_s\rangle_A = \frac{1}{\sqrt{2}} ( |1,0\rangle_s + e^{i\theta_s} |0,1\rangle_s )$. The action of the local projection on the joint space is $\hat{\Pi}_A(\theta_s) \otimes \hat{I}_B = |\theta_s\rangle_A ( {}_A\langle \theta_s| \otimes \hat{I}_B )$. Applying the linear functional ${}_A\langle \theta_s| \otimes \hat{I}_B$ directly to the formal components of $[v_{AB}]$ gives:
\begin{equation}
\left( {}_A\langle \theta_s| \otimes \hat{I}_B \right) [v_{AB}] = c_1 \, {}_A\langle \theta_s | 1,0\rangle_s |m_1\rangle_i + c_2 \, {}_A\langle \theta_s | 0,1\rangle_s |m_2\rangle_i.
\label{eq:app_contraction}
\end{equation}
The inner products of the detector bra with the path states evaluate to:
\begin{align}
{}_A\langle \theta_s | 1,0\rangle_s &= \frac{1}{\sqrt{2}} \left( {}_s\langle 1,0| + e^{-i\theta_s} {}_s\langle 0,1| \right) |1,0\rangle_s = \frac{1}{\sqrt{2}}, \label{eq:app_inner1} \\
{}_A\langle \theta_s | 0,1\rangle_s &= \frac{1}{\sqrt{2}} \left( {}_s\langle 1,0| + e^{-i\theta_s} {}_s\langle 0,1| \right) |0,1\rangle_s = \frac{e^{-i\theta_s}}{\sqrt{2}}. \label{eq:app_inner2}
\end{align}
Substituting Eqs.~(\ref{eq:app_inner1}) and (\ref{eq:app_inner2}) into Eq.~(\ref{eq:app_contraction}) yields:
\begin{equation}
\left( {}_A\langle \theta_s| \otimes \hat{I}_B \right) [v_{AB}] = \frac{1}{\sqrt{2}} \left( c_1 |m_1\rangle_i + c_2 e^{-i\theta_s} |m_2\rangle_i \right) \equiv |\chi_B(\theta_s)\rangle.
\label{eq:app_coset_state}
\end{equation}
Recombining with the detector state $|\theta_s\rangle_A$, the projected coset slice is $( \hat{\Pi}_A(\theta_s) \otimes \hat{I}_B ) [v_{AB}] = |\theta_s\rangle_A \otimes |\chi_B(\theta_s)\rangle$, with $|\chi_B(\theta_s)\rangle = \frac{1}{\sqrt{2}} ( c_1 |m_1\rangle_i + c_2 e^{-i\theta_s} |m_2\rangle_i )$, which proves Eq.~(\ref{eq:chi_B_secB}).

\section{Operational Measurement Physics: Dual-Bra Contraction and Three-Scan Phase Equivalence}
\label{app:measurement}
\label{app:measurement_physics}

In this appendix, we provide a detailed operational analysis of the measurement process on the composite entangled state, explicitly deriving the relative phase factor $e^{-i\theta_s}$ appearing in the conditional idler state $|\chi_B(\theta_s)\rangle$ [Eq.~(\ref{eq:chi_B_secB})] and proving the physical equivalence among the pump, seed, and signal phase scans.

Before measurement, the bipartite state generated across the dual PPLN crystals is represented by the equivalence class in the quotient tensor product space $\mathcal{H}_{AB} \cong \mathcal{H}_s \otimes \mathcal{H}_i$:
\begin{equation}
[v_{AB}] = c_1 |1,0\rangle_s \otimes |m_1\rangle_i + c_2 |0,1\rangle_s \otimes |m_2\rangle_i,
\label{eq:appC_state}
\end{equation}
where the path-encoded signal Fock states $|1,0\rangle_s$ and $|0,1\rangle_s$ are strictly correlated with the bipartite idler states $|m_1\rangle_i$ and $|m_2\rangle_i$, respectively.

At the output of the signal Mach--Zehnder interferometer, the two spatial paths combine at a 50:50 beam splitter (BS3) prior to photodetection. A single-photon registration at the detector $\mathrm{PD}$ corresponds to the projection onto a normalized superposition state $|\theta_s\rangle_A = \frac{1}{\sqrt{2}} ( |1,0\rangle_s + e^{i\theta_s} |0,1\rangle_s ) \in \mathcal{H}_s$, where the interferometric phase $\theta_s = \Delta\phi_s = k_s \Delta x_s$ is controlled by displacing the mirror mounted on a piezoelectric transducer (PZT). 

In quantum measurement theory, the projection operator associated with this detection event is the rank-1 operator $\hat{\Pi}_A(\theta_s) = |\theta_s\rangle_A \langle \theta_s|_A$. The linear functional performing the measurement is the dual bra vector ${}_A\langle \theta_s|$, obtained via Hermitian conjugation:
\begin{equation}
{}_A\langle \theta_s| = \left( |\theta_s\rangle_A \right)^\dagger = \frac{1}{\sqrt{2}} \left( {}_s\langle 1,0| + e^{-i\theta_s} {}_s\langle 0,1| \right).
\label{eq:appC_bra}
\end{equation}
Crucially, complex conjugation inverts the sign of the relative phase from $+i\theta_s$ in the state ket to $-i\theta_s$ in the dual bra.

Acting with the joint measurement operator $\hat{\Pi}_A(\theta_s) \otimes \hat{I}_B = |\theta_s\rangle_A ( {}_A\langle \theta_s| \otimes \hat{I}_B )$ upon the entangled state $[v_{AB}]$, the linear map operates on each path component independently: $( {}_A\langle \theta_s| \otimes \hat{I}_B ) [v_{AB}] = c_1 ( {}_A\langle \theta_s | 1,0\rangle_s ) |m_1\rangle_i + c_2 ( {}_A\langle \theta_s | 0,1\rangle_s ) |m_2\rangle_i$. Because the path basis states $\{|1,0\rangle_s, |0,1\rangle_s\}$ are orthonormal, the inner products evaluate to ${}_A\langle \theta_s | 1,0\rangle_s = 1/\sqrt{2}$ and ${}_A\langle \theta_s | 0,1\rangle_s = e^{-i\theta_s}/\sqrt{2}$. The state of the unmeasured idler field conditioned upon registering the signal photon in mode $|\theta_s\rangle_A$ is therefore:
\begin{equation}
|\chi_B(\theta_s)\rangle = \left( {}_A\langle \theta_s| \otimes \hat{I}_B \right) [v_{AB}] = \frac{1}{\sqrt{2}} \left( c_1 |m_1\rangle_i + c_2 e^{-i\theta_s} |m_2\rangle_i \right).
\label{eq:appC_chi_result}
\end{equation}
This confirms that the phase factor $e^{-i\theta_s}$ does not originate from an active physical phase shift or non-local action applied to the idler beam. Rather, it is the coordinate weight of the local signal projection operator projected into the dual basis of the measurement apparatus.

The operational detection probability (photocount rate $R_D$) registered at the signal detector is proportional to the norm square of the conditional coset state: $R_D \propto \langle \chi_B(\theta_s) | \chi_B(\theta_s) \rangle$. Expanding this inner product:
\begin{align}
\langle \chi_B(\theta_s) | \chi_B(\theta_s) \rangle &= \frac{1}{2} \Big[ |c_1|^2 \langle m_1|m_1\rangle_i + |c_2|^2 \langle m_2|m_2\rangle_i \notag \\
&\quad + c_1 c_2^* \langle m_2|m_1\rangle_i e^{i\theta_s} + c_1^* c_2 \langle m_1|m_2\rangle_i e^{-i\theta_s} \Big].
\label{eq:appC_norm_expanded}
\end{align}
Using the normalization condition $\langle m_j|m_j\rangle_i = 1$ and $|c_1|^2 + |c_2|^2 = 1$, together with the idler overlap $\langle m_2 | m_1 \rangle_i = F \exp[ -i ( \Delta\phi_p - \Delta\phi_{\text{seed}} ) ]$, the interference cross-term becomes:
\begin{equation}
c_1 c_2^* \langle m_2|m_1\rangle_i e^{i\theta_s} = |c_1 c_2| F \exp\left[ i \left( \Delta\phi_s - \Delta\phi_p + \Delta\phi_{\text{seed}} \right) \right],
\label{eq:appC_cross_term}
\end{equation}
where $\theta_s = \Delta\phi_s$. Substituting Eq.~(\ref{eq:appC_cross_term}) into Eq.~(\ref{eq:appC_norm_expanded}), the single-photon count rate simplifies to:
\begin{equation}
R_D(\Delta\Phi) = \frac{R_0}{2} \left[ 1 + V \cos\left( \Delta\Phi \right) \right],
\label{eq:appC_fringe_law}
\end{equation}
where $V = 2|c_1 c_2| F = V_0 F$ is the interferometric visibility, and the total operational phase $\Delta\Phi$ is given by:
\begin{equation}
\Delta\Phi = \Delta\phi_s - \Delta\phi_p + \Delta\phi_{\text{seed}}.
\label{eq:appC_total_phase}
\end{equation}

Equation~(\ref{eq:appC_total_phase}) rigorously explains the experimental observation in Ref.~\cite{Lee2018}: scanning the signal arm difference $\Delta x_s$ varies $\Delta\phi_s = 2\pi \Delta x_s / \lambda_s$, producing sinusoidal fringes with spatial period $\lambda_s \approx 807.2\text{ nm}$; scanning the pump comb path difference $\Delta x_p$ varies $\Delta\phi_p = 2\pi \Delta x_p / \lambda_p$, producing identical sinusoidal fringes with spatial period $\lambda_p = 530\text{ nm}$; and scanning the seed arm difference $\Delta x_{\text{seed}}$ varies $\Delta\phi_{\text{seed}} = 2\pi \Delta x_{\text{seed}} / \lambda_i$, producing identical sinusoidal fringes with spatial period $\lambda_i = 1542.4\text{ nm}$. As established by Yoon, all three scans modulate the single invariant scalar phase $\Delta\Phi$ governing the projection of the composite state onto the detector-selected subspace, proving that single-particle interference is fundamentally defined by the joint relation between the source state and the measurement basis rather than an isolated local path accumulation.

\section{Relativistic QED Formulation and Microcausality of Transverse Field Operators}
\label{app:qed}
\label{app:rel_qed_derivation}

To establish the strict relativistic microcausality of the dual-crystal single-photon frequency-comb (SPFC) architecture, we formulate the electromagnetic field within covariant relativistic quantum electrodynamics (QED) using the Coulomb (transverse) gauge. In this gauge, the transverse vector potential operator $\hat{\mathbf{A}}_\perp(x)$ at spacetime point $x = (ct, \mathbf{x})$ satisfies the wave equation $\Box \hat{\mathbf{A}}_\perp(x) = 0$ and is expanded in terms of plane-wave modes and multi-mode comb frequency envelopes:

\begin{equation}
\hat{\mathbf{A}}_\perp(x) = \sum_{\mathbf{k}, \lambda} \sqrt{\frac{\hbar}{2\epsilon_0 V \omega_{\mathbf{k}}}} \left[ \mathbf{e}_{\mathbf{k},\lambda} \hat{a}_{\mathbf{k},\lambda} e^{-i k \cdot x} + \mathbf{e}_{\mathbf{k},\lambda}^* \hat{a}_{\mathbf{k},\lambda}^\dagger e^{i k \cdot x} \right],
\label{eq:app_transverse_potential}
\end{equation}
where $k \cdot x = \omega_{\mathbf{k}} t - \mathbf{k} \cdot \mathbf{x}$, $\mathbf{e}_{\mathbf{k},\lambda}$ are the transverse polarization vectors ($\mathbf{k} \cdot \mathbf{e}_{\mathbf{k},\lambda} = 0$), and $\hat{a}_{\mathbf{k},\lambda}^{(\dagger)}$ are the photon annihilation (creation) operators satisfying the canonical commutation relations $[\hat{a}_{\mathbf{k},\lambda}, \hat{a}_{\mathbf{k}',\lambda'}^\dagger] = \delta_{\mathbf{k},\mathbf{k}'}\delta_{\lambda,\lambda'}$.

To verify that local measurements performed at station $A$ do not cause superluminal disturbances at station $B$ across the laboratory baseline $L_0 \approx 1\text{ m}$, we evaluate the equal-time and spacelike commutation relations of the transverse vector potential operators. The commutator between transverse field operators evaluated at two distinct spacetime points $x_1 = (ct_1, \mathbf{x}_1)$ and $x_2 = (ct_2, \mathbf{x}_2)$ is given by the invariant Jordan-Pauli function:
\begin{equation}
[\hat{\mathbf{A}}_\perp(x_1), \hat{\mathbf{A}}_\perp(x_2)] = i \hbar D_\perp(x_1 - x_2) \mathbf{I},
\label{eq:app_commutator}
\end{equation}
where $D_\perp(x_1 - x_2)$ represents the transverse components of the invariant commutator (Jordan-Pauli) function evaluated in the transverse gauge. For any two measurement events separated by a spacelike interval where $(x_1 - x_2)^2 = c^2(t_1 - t_2)^2 - |\mathbf{x}_1 - \mathbf{x}_2|^2 < 0$, the commutator vanishes identically:
\begin{equation}
[\hat{\mathbf{A}}_\perp(x_1), \hat{\mathbf{A}}_\perp(x_2)] = 0.
\label{eq:app_microcausality}
\end{equation}

Equation~(\ref{eq:app_microcausality}) guarantees strict relativistic microcausality. Consequently, when an observer at station $A$ measures the signal field using the local projection operator $\hat{\Pi}_A(\theta_s)$, the emergence of the conditional coset slice $|\chi_B(\theta_s)\rangle$ at station $B$ is not mediated by a physical superluminal wave-packet collapse propagating through the intervening space. Instead, as proved in the quotient-space formulation, the correlation is pre-encoded in the invariant equivalence class geometry $[v_{AB}]$ and is unmasked via linear algebraic contraction, preserving relativistic causality at all spacetime coordinates.

\bigskip
Tai Hyun Yoon 0000-0002-2408-9295

\clearpage
\onecolumngrid
\begin{center}
{\large\bfseries Supplemental Material}\\[4pt]
{\itshape Resolving nonlocal state reduction via induced coherence in frequency-comb biphoton sources}\\[2pt]
Tai Hyun Yoon
\end{center}

\setcounter{section}{0}
\renewcommand{\thesection}{S\arabic{section}}
\renewcommand{\thesubsection}{S\arabic{section}\Alph{subsection}}
\setcounter{equation}{0}
\renewcommand{\theequation}{S\arabic{equation}}
\setcounter{figure}{0}
\renewcommand{\thefigure}{S\arabic{figure}}
\setcounter{table}{0}
\renewcommand{\thetable}{S\arabic{table}}
\section{Detailed Experimental Architecture and Gaussian Beam Interaction Length}
\label{sec:S1}
\label{sec:experimental_setup}

The dual-crystal single-photon frequency-comb (SPFC) architecture (see Fig.~\ref{fig:fig1}) is designed to satisfy strict relativistic microcausality across a macroscopic laboratory baseline. The two sources are separated by a spatial distance of $L_0 \approx 1\text{ m}$ \cite{Lee2018}. 

The pump beam is generated by an optical frequency comb at 1060~nm (46~nm spectral bandwidth) which is frequency-doubled using an LBO crystal. The resulting 530~nm pump comb ($\sim$3~nm spectral width, $\sim$14~mW average power) is phase-locked to a repetition rate of $f_r = 250.0$~MHz and a carrier-offset frequency of $f_{\text{ceo}} = 20.0$~MHz, referenced to a GPS-disciplined Rb atomic clock \cite{Lee2018}. 

The nonlinear sources are identically periodically poled lithium niobate (PPLN) crystals with type-0 phase matching, physical dimensions of $15 \times 7.9 \times 0.5~\text{mm}^3$, and a grating period of 7.3~$\mu$m. They are maintained at a phase-matching temperature of 121.5~$^\circ$C. The unmeasured conjugate environment is established via an ultra-narrow continuous-wave (cw) seed laser injected along the pump paths using dichroic mirrors. The seed wavelength is 1542.384~nm with a linewidth of $\Delta\nu < 10$~Hz \cite{Lee2018}.

While the physical length of each PPLN crystal is 15~mm, the effective active interaction length $L_c$ is governed by the confocal parameter $b$ of the focused pump beam. The input pump and seed beams, with collimated diameters of approximately 2~mm, are focused into the crystals using lenses with a focal length of $f = 100$~mm. For the 530~nm pump, this geometry produces a beam waist radius of $w_0 \approx 17.3~\mu$m inside the crystal. Using the extraordinary refractive index $n \approx 2.23$, the resulting confocal parameter is $b = 2\pi w_0^2 n / \lambda_p \approx 7.9$~mm. Consequently, the effective spatial interaction length driving the spatial Lindblad master equation is strictly bounded to $L_c \approx 7.9$~mm \cite{Lee2018}.

The unheralded signal single-photon supermodes at 807.2~nm (bandwidth $> 0.4$~nm) are recombined on a beam splitter and detected using an EMCCD-coupled spectrometer (0.1~nm resolution) \cite{Lee2018}. To map the single-particle interference fringes across the triality parameter space, the detection apparatus operates with a 10-ms integration window. The ensemble interference fringes are averaged over $2.5 \times 10^6$ pulses at a deep single-photon filling factor of approximately 0.02 photons per pulse \cite{Lee2018}.

\section{Numerical and Computational Methodologies}
\label{sec:S2}
\label{sec:numerical_methods}

\begin{figure}[htbp]
    \centering
    \includegraphics[width=0.95\textwidth]{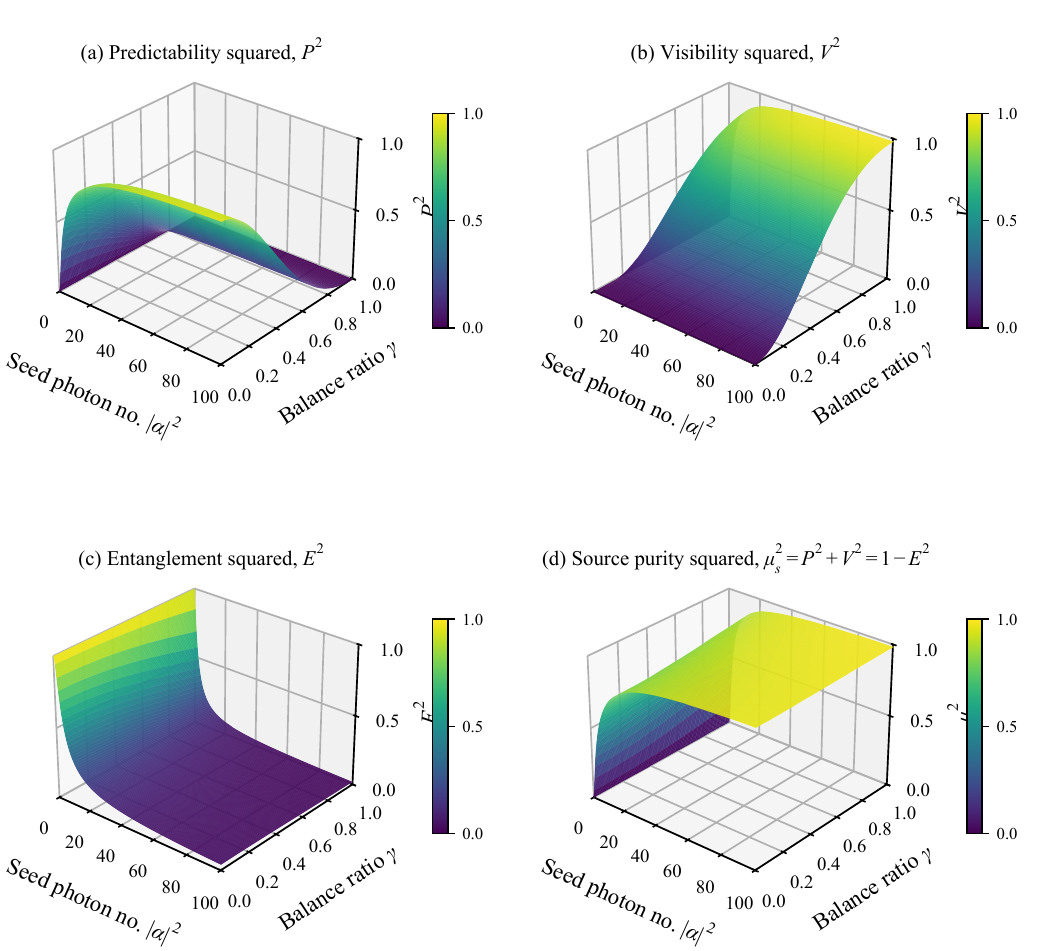}
    \caption{\textbf{Theoretical parameter space mapping of the quantitative wave-particle-entanglement triality relation.} The continuous 3D surfaces represent the exact analytical solutions derived from the spatial Lindblad open-system framework for (a) predictability squared $P^2$, (b) visibility squared $V^2$, (c) entanglement squared $E^2$, and (d) source purity squared $\mu_s^2$. The physical observables are plotted as a function of the local conjugate seed parameters: the seed photon number $|\alpha|^2 \in [0, 100]$ and the balance ratio $\gamma \in [0, 1]$. These computational surfaces visually confirm the strict saturation of the triality equality $P^2 + E^2 + V^2 = 1$ and the source purity relation $\mu_s^2 = P^2 + V^2 = 1 - E^2$ across the complete measurement domain, demonstrating that which-path distinguishability is fundamentally partitioned between classical beam imbalance and quantum entanglement with the unobserved environment.}
    \label{fig:figS1}
\end{figure}

The analytical solutions for the open-system spatial Lindblad master equation were verified using rigorous numerical simulations. Matrix exponentiation and the spatial integration of the reduced density matrix $\hat{\rho}_s(y)$ were performed using Mathematica. 

The effective spatial dephasing rate, $\Gamma_{\text{deph}} = -(1/L_c) \ln F$, is governed by the state overlap fidelity $F$ of the single-photon-added coherent states (SPACS). The fidelity is continuously controlled experimentally via the transmission ratio $\gamma$ and the seed photon number $|\alpha|^2$ \cite{Yoon2021}:
\begin{equation}
    F = \frac{\gamma |\alpha|^2}{\sqrt{(1+|\alpha|^2)(1+\gamma^2|\alpha|^2)}}.
    \label{eq:SM_SPACS_fidelity}
\end{equation}

To map the wave-particle-entanglement triality relation $P^2 + E^2 + V^2 = 1$ across the complete two-dimensional $\{\gamma, |\alpha|^2\}$ parameter space, the explicit analytical formulas for predictability squared ($P^2$), visibility squared ($V^2$), and entanglement squared ($E^2$) are derived as \cite{Yoon2021}:
\begin{align}
    P^2 &= \left[ \frac{|\alpha|^2(1-\gamma^2)}{2+|\alpha|^2(1+\gamma^2)} \right]^2, \label{eq:SM_P2} \\
    V^2 &= \left[ \frac{2\gamma|\alpha|^2}{2+|\alpha|^2(1+\gamma^2)} \right]^2, \label{eq:SM_V2} \\
    E^2 &= \frac{4[1+|\alpha|^2(1+\gamma^2)]}{[2+|\alpha|^2(1+\gamma^2)]^2}. \label{eq:SM_E2}
\end{align}
The source purity squared is identically constrained by these observables via $\mu_s^2 = P^2 + V^2 = 1 - E^2$ \cite{Yoon2021}.

Using Python (\texttt{matplotlib} and \texttt{numpy}), these exact algebraic relations were computationally evaluated to generate the continuous 3D surface manifolds shown in Fig.~\ref{fig:figS1}. These theoretical plots visually confirm the strict saturation of the triality equality across the entire physical measurement domain, demonstrating that which-path distinguishability is partitioned strictly between classical beam imbalance and quantum entanglement with the unobserved idler reservoir \cite{Yoon2021}.

\section{Hardware-Native TF-GKP Qubit Parameters and Error-Correction Thresholds}
\label{sec:S3}
\label{sec:tfgkp_parameters}

The macroscopic single-photon frequency-comb (SPFC) supermode generated in this architecture is a finite-energy codeword of a hardware-native continuous-variable time-frequency Gottesman-Kitaev-Preskill (TF-GKP) qubit \cite{Yoon2026a}. Unlike standard discrete-variable encodings, the TF-GKP formalism embeds a logical qubit within the infinite-dimensional Hilbert space spanned by the continuous time ($\hat{T}$) and frequency ($\hat{\Omega}$) modal quadratures of a single photon, which obey the canonical commutation relation $[\hat{T}, \hat{\Omega}] = i$ \cite{Gottesman2001, Yoon2026a}.

In our physical realization, the ideal, unphysical GKP grid of Dirac delta peaks is naturally regularized by the finite temporal and spectral bounds of the interacting fields. The SPFC state $|\Psi_{\text{SPFC}}\rangle = \hat{S}_j^\dagger|0\rangle$ encompasses $N \approx 209$ phase-locked comb teeth within the phase-matching bandwidth of the PPLN crystal. The overall finite-energy spectral envelope is governed by the $8.4$-ps duration of the pump pulses, while the macroscopic temporal envelope (coherence time) is determined by the narrow ($<10$~Hz) linewidth of the continuous-wave seed laser \cite{Yoon2026a}. Its teeth lie on the grid $\Omega_0 + n\omega_r$, with $\Omega_0 = 2\pi f_{\text{ceo}} - \omega_{\text{seed}} \pmod{\omega_r}$ the fractional offset fixed by the pump carrier-envelope offset and the seed frequency.

Because the comb structure is intrinsically phase-locked to a GPS-disciplined Rb atomic clock, the TF-GKP lattice is deterministically anchored at the hardware layer. The two displacements that leave the SPFC comb invariant,
\begin{align}
    \hat{\mathcal{Q}}_{\tau} &= \exp[-i\Delta T (\hat{\Omega}-\Omega_0)], \label{eq:SM_Qtau} \\
    \hat{\mathcal{Q}}_{\omega} &= \exp(i\omega_r \hat{T}), \label{eq:SM_Qomega}
\end{align}
commute because $\Delta T\,\omega_r = 2\pi$, but their primitive cell has symplectic area $2\pi$ and supports only a one-dimensional common eigenspace: Eqs.~(\ref{eq:SM_Qtau})--(\ref{eq:SM_Qomega}) are the stabilizers of a GKP \emph{qunaught} (sensor) lattice \cite{Walshe2020}, not of a qubit code. A logical qubit requires a stabilizer cell of area $4\pi$. With the logical translation scales $T_L = \Delta T$ and $\Omega_L = \omega_r/2$ ($T_L\Omega_L = \pi$), the logical Pauli translations and the stabilizers of the TF-GKP code are
\begin{align}
    \bar{Z} &= \exp[-iT_L(\hat{\Omega}-\Omega_0)] = \hat{\mathcal{Q}}_\tau, \qquad \bar{X} = \exp(i\Omega_L\hat{T}), \label{eq:SM_logicals} \\
    \hat{\mathcal{S}}_{T} &= \bar{Z}^2 = \exp[-2i\Delta T (\hat{\Omega}-\Omega_0)], \label{eq:SM_Stau} \\
    \hat{\mathcal{S}}_{\Omega} &= \bar{X}^2 = \exp(i\omega_r \hat{T}) = \hat{\mathcal{Q}}_\omega, \label{eq:SM_Somega}
\end{align}
with $\bar{X}\bar{Z} = -\bar{Z}\bar{X}$ and $[\hat{\mathcal{S}}_{T}, \hat{\mathcal{S}}_{\Omega}] = 0$. The logical code space is the common $+1$ eigenspace of $\hat{\mathcal{S}}_T$ and $\hat{\mathcal{S}}_\Omega$; its codewords are the two cosets of the interleaved grid $\Omega_0 + \mu\,\omega_r/2$, namely $|\bar{0}\rangle \propto \sum_n |\Omega_0 + n\omega_r\rangle$ (the SPFC supermode as generated) and $|\bar{1}\rangle \propto \sum_n |\Omega_0 + (n+\tfrac12)\omega_r\rangle$ (the same comb with the seed, or the pump carrier-envelope offset, shifted by $f_r/2$). In the time domain, $|\bar{0}\rangle$ and $|\bar{1}\rangle$ are pulse trains of period $\Delta T$ whose carrier phase slips by $2\pi\Omega_0/\omega_r$ or by $2\pi\Omega_0/\omega_r + \pi$ per round trip, and $|\bar{\pm}\rangle$ occupy only even or only odd time slots. The codeword $|\bar{0}\rangle$ is at the same time the qunaught state of the $2\pi$ lattice of Eqs.~(\ref{eq:SM_Qtau})--(\ref{eq:SM_Qomega}), whereas $|\bar{+}\rangle$, the comb on the half-integer grid, is the qunaught state of the dual lattice $(2\Delta T, \omega_r/2)$. For our system, the fundamental grid spacing is fixed by the pump comb repetition rate $f_r = 250.0$ MHz, yielding a temporal bin size (pulse repetition interval) of $\Delta T = 1/f_r = 4.0$ ns and a spectral tooth spacing of $\omega_r = 2\pi f_r \approx 2\pi \times 250.0$ MHz \cite{Yoon2026a}; the logical half-tooth shift is therefore $f_r/2 = 125.0$~MHz, well within the range of an acousto-optic modulator on the seed beam.

A bichromatic seed with components at $\omega_{\text{seed}}$ and $\omega_{\text{seed}} + \omega_r/2$ in a single crystal prepares $c_1|\bar{0}\rangle|m_1\rangle_i + c_2|\bar{1}\rangle|m_2\rangle_i$, with the coefficients of Eq.~(\ref{eq:norm_coeff}) and the single-photon-added coherent idler states of Eq.~(\ref{eq:m_states}) attached to the two seed components. The seed balance $\gamma$ sets the logical amplitudes and the relative seed phase sets the logical phase, so one crystal steers the conditional single-photon state over the full logical Bloch sphere: $\gamma = 1$ with zero relative phase gives $|\bar{+}\rangle$, and a relative phase of $\pi/4$ gives the magic state $(|\bar{0}\rangle + e^{i\pi/4}|\bar{1}\rangle)/\sqrt{2}$. The purity of the encoded qubit is governed by the SPACS overlap $F$ of Eq.~(\ref{eq:SM_SPACS_fidelity}), and the surfaces of Fig.~\ref{fig:figS1} can be read directly as the logical predictability, logical coherence, and idler entanglement of this qubit. These statements refer to the one-photon component of the seeded output conditioned on the registration of one signal photon. The unconditioned stimulated field is a displaced thermal state with Poissonian statistics for $|\alpha|^2 \gg 1$ ($P_2/P_1 = \bar{n}/2$ per time bin), and the idler modes carry only the information $E^2 = 1 - F^2$ about an emission, so the seeded carrier cannot be heralded; the seeded source is a bright, phase-coherent twin of the qubit mode for alignment, calibration, and logical tomography, not a single-photon qubit source.

Environmental decoherence manifests as small temporal displacements (timing jitter, $e^{-i\delta t \hat{\Omega}}$) and spectral drifts (frequency offsets, $e^{i\delta\omega \hat{T}}$). The TF-GKP code tolerates these continuous shifts passively, through the modulo structure of its logical readout (slot parity for $\bar{X}$, tooth parity for $\bar{Z}$), provided the errors remain within the Voronoi cell of the code:
\begin{align}
    |\delta t| &< \frac{\Delta T}{2} = 2.0~\text{ns}, \label{eq:SM_jitter_bound} \\
    |\delta \omega| &< \frac{\omega_r}{4} = 2\pi \times 62.5~\text{MHz}. \label{eq:SM_drift_bound}
\end{align}
The spectral bound is half of the qunaught cell, because a shift by $\omega_r/2$ is the logical operator $\bar{X}$ rather than a correctable displacement. Our dual-PPLN SPFC architecture fulfills these criteria completely at the native hardware level without requiring auxiliary Gaussian squeezing interactions. The $8.4$-ps pump pulse duration ensures that the temporal localization of each comb tooth is deeply sub-nanosecond, thereby securely satisfying $|\delta t| \ll 2.0$ ns. Concurrently, the Hertz-level linewidth of the individual comb teeth ($\sigma_\Omega \approx 10^{-8}\,\omega_r$), guaranteed by the cw seed laser and the Rb-disciplined OFC, ensures that spectral broadening is negligible, strictly satisfying $|\delta\omega| \ll \omega_r/4$ \cite{Yoon2026a}. Consequently, the SPFC encoding provides a displacement-tolerant physical qubit for measurement-based and fusion-based photonic quantum computation. Three limits of this statement should be kept explicit: the encoding does not correct photon loss; it does not supply a deterministic entangling gate between photons; and the seeded source, being unheraldable, does not supply the single photons such a computation consumes. Heralded single photons and the entangling resource both come from the signal--idler pair emitted by the same crystal in the unseeded regime, whose frequencies satisfy $\omega_s + \omega_i = \omega_{p,n}$: in a bare crystal this pair is the collective two-photon TF-GKP code \cite{Descamps2024} (a comb in the sum frequency, each photon alone being diagonal in its teeth), and when the crystal is placed in a comb-locked cavity of free spectral range $f_r/2$, which confines each photon to the interleaved grid, it becomes a Bell pair of single-mode TF-GKP qubits with parity set by the pump carrier-envelope offset; loss tolerance and two-qubit gates are then supplied by linear-optical fusion in an outer code \cite{Bartolucci2023}.

\section{Rigorous Quotient Measurement Theory: Algebraic, Geometric, and Operational Equivalence}
\label{sec:S4}
\label{sec:measurement_terminology}

To resolve the paradox of nonlocal state reduction without invoking superluminal dynamical collapse, informal geometric intuitions—such as extracting a ``slice'' of a wavefunction—are replaced with exact mathematical descriptions \cite{Scholes2026, Yoon2026b}. In our framework, local measurement of the signal photon across a spacelike-separated baseline determines the properties of the unmeasured idler field through three equivalent perspectives: algebraic, geometric, and operational. 

These formalisms describe the identical physical process, entirely bypassing the projection postulate's requirement for a remote physical update \cite{Brunner2014, Scholes2026}.

\subsection{The Algebraic Perspective: Conditional Cosets}
In the free-vector-space quotient construction, the bipartite state prior to measurement is not a rigid geometric object, but an equivalence class $[v_{AB}]$ residing in the quotient space $\mathcal{H}_{AB} = \mathcal{F}(V_A \times V_B)/\mathcal{R}$ \cite{Scholes2026}. 
\begin{itemize}
    \item Rather than physically cutting a ``slice'' out of a collapsed wavefunction, a local measurement corresponds to evaluating a linear functional (the dual bra ${}_A\langle\theta_s| \otimes \hat{I}_B$) directly on the formal sum of the equivalence class \cite{Scholes2026, Yoon2026b}.
    \item This algebraic contraction mathematically isolates the \textbf{conditional coset} $|\chi_B(\theta_s)\rangle = \frac{1}{\sqrt{2}}(c_1|m_1\rangle_i + c_2 e^{-i\theta_s}|m_2\rangle_i)$ representing the unmeasured idler field \cite{Scholes2026, Yoon2026b}. 
    \item Because the correlation is pre-encoded in the contextual phase structure of the equivalence class, the conditional coset emerges naturally from the local algebra without any superluminal action propagating from region $A$ to region $B$ \cite{Brunner2014, Scholes2026}.
\end{itemize}

\subsection{The Geometric Perspective: Local Subspaces}
From the standpoint of standard Hilbert space geometry, the measurement apparatus defines a rank-1 projection operator $\hat{\Pi}_A(\theta_s) = |\theta_s\rangle_A \langle\theta_s|_A$ acting strictly within the local carrier algebra $\mathcal{B}(V_A)$ \cite{Scholes2026}:
\begin{itemize}
    \item When the signal photon is registered, the composite state is projected onto a specific \textbf{local subspace} of the signal's vector space $V_A$, spanned by the ray $|\theta_s\rangle_A$ \cite{Scholes2026}.
    \item This geometric restriction in $V_A$ correlates with the state of the idler environment in $V_B$ via the bipartite entanglement structure, meaning the nonlocal correlation is a consequence of local subspace filtering rather than an active modification of the distant field \cite{Brunner2014, Scholes2026}. 
    \item The operational single-photon counting rate is simply the geometric trace (the Hilbert-Schmidt norm) evaluated over this localized subspace \cite{Lee2018, Yoon2026b}.
\end{itemize}

\subsection{The Operational Perspective: Local Measurement Bases}
In the laboratory, the abstract subspace is physically instantiated by the interferometer hardware:
\begin{itemize}
    \item The term \textbf{local measurement basis} defines the specific coordinate frame selected by the experimentalist, corresponding to the state vector $|\theta_s\rangle_A = \frac{1}{\sqrt{2}}(|1,0\rangle_s + e^{i\theta_s}|0,1\rangle_s)$ \cite{Yoon2026b}.
    \item Shifting the macroscopic mirror via the piezoelectric transducer (PZT) modulates the signal interferometric readout phase $\theta_s \equiv \phi_D \equiv \Delta\phi_s$, effectively rotating the local measurement basis \cite{Lee2018, Yoon2026b}.
    \item As established by the three-scan phase equivalence, single-particle interference is fundamentally dictated by the relative phase between the prepared composite quantum state and this tunable, detector-defined local measurement basis \cite{Yoon2026b}.
\end{itemize}

In summary, the macroscopic appearance of state reduction is completely described by the projection onto a \textit{local subspace} governed by the \textit{local measurement basis}, which algebraically extracts the corresponding \textit{conditional coset} \cite{Scholes2026, Yoon2026b}. This tri-fold equivalence eliminates the need for \textit{ad hoc} wave mechanics modifications and preserves strict relativistic microcausality across the 1-m PPLN baseline \cite{Brunner2014, Lee2018}.

\section{QED Gauge Symmetry of Bilinear Relations $\mathcal{R}$ and One-to-One Correspondence with Contextual-Phase Formalisms}
\label{sec:S5}
\label{sec:gauge_and_correspondence}

In two recent theoretical treatises, Scholes~\cite{Scholes2026a,Scholes2026} addressed the century-long dilemma of quantum measurement by proposing that entangled states are equivalence classes $[\Psi] \in \mathcal{F}/\mathcal{R}$ in a covering free vector space whose internal ``contextual phase'' directs collapse upon local projection. In this section, we establish the systematic one-to-one correspondence between that algebraic framework and our physical, frequency-comb biphoton architecture.

\subsection{Subspace of Relations $\mathcal{R}$: Algebraic Equivalence vs.\ Physical QED Gauge Symmetry}
\label{subsec:SM_gauge_symmetry}

In Scholes (Sec.~4 in~\cite{Scholes2026a}; Sec.~4.2, Eq.~3 in~\cite{Scholes2026}), $\mathcal{R} \subset \mathcal{F}(\mathcal{H}_A \times \mathcal{H}_B)$ is defined algebraically as the linear subspace generated by formal bilinear differences:
\begin{align}
    (r u_1 + s u_2, v) - r(u_1, v) - s(u_2, v) &\in \mathcal{R}, \label{eq:SM_R_lin1} \\
    (u, r v_1 + s v_2) - r(u, v_1) - s(u, v_2) &\in \mathcal{R}, \label{eq:SM_R_lin2}
\end{align}
with $r, s \in \mathbb{C}$. In particular, setting $r = e^{i\phi}$ generates the scalar equivalence relation:
\begin{equation}
    (e^{i\phi} u, v) - (u, e^{i\phi} v) \in \mathcal{R},
    \label{eq:SM_scholes_R_scalar}
\end{equation}
which identifies formal basis points $(e^{i\phi} u, v) \sim (u, e^{i\phi} v)$ modulo $\mathcal{R}$. In Scholes's framework, $\mathcal{R}$ is an abstract algebraic construct ensuring the bilinearity of the tensor product, which partitions points in $\mathbb{S}_A^3 \times \mathbb{S}_B^3$ into an equivalence class $[\Psi] \in \mathcal{F}/\mathcal{R}$.

\paragraph{Physical Realization of $\mathcal{R}$ in Our Architecture:}
In our macroscopic frequency-comb biphoton platform, the relation subspace $\mathcal{R}$ represents the \textbf{physical phase-matching conditions, local $U(1)$ gauge invariance in QED three-wave mixing, and spatial interferometric recombination}:
\begin{enumerate}
    \item \textbf{Microscopic QED Origin and $U(1)$ Phase Invariance:} 
    At the fundamental field level, the second-order nonlinear susceptibility $\chi^{(2)}$ originates from closed electronic triangle loops governed by the local $U(1)$ gauge-invariant interaction density $\mathcal{L}_{\text{int}} = -j_\mu A^\mu$, where $j_\mu = q\bar{\psi}\gamma_\mu\psi$ is the conserved Dirac four-current density of the crystal electrons and $A^\mu$ is the electromagnetic four-potential. In the spatial propagation picture, three-wave mixing is governed by the effective Hamiltonian:
    \begin{equation}
        \hat{H}_{\text{int}}(y) \propto \chi^{(2)} \hat{a}_p(y) \hat{a}_s^\dagger(y) \hat{a}_i^\dagger(y) e^{i\Delta k y} + \text{H.c.},
    \end{equation}
    where spatial translational quasi-symmetry enforces phase matching $\theta_p = \theta_s + \theta_i \pmod{2\pi}$. Under the internal gauge transformation $\hat{a}_s \to \hat{a}_s e^{i\lambda}$ and $\hat{a}_i \to \hat{a}_i e^{-i\lambda}$, the Hamiltonian $\hat{H}_{\text{int}}(y)$ is strictly invariant. Consequently, the scalar equivalence relations in Eq.~(\ref{eq:SM_scholes_R_scalar}) physically define the internal redistribution of phase along this continuous \textbf{gauge orbit}:
    \begin{equation}
        \big(|1,0\rangle_s, e^{i(\Delta\phi_p - \Delta\phi_{\text{seed}})} |v_2\rangle_i\big) - \big(e^{-i(\Delta\phi_p - \Delta\phi_{\text{seed}})} |1,0\rangle_s, |v_2\rangle_i\big) \in \mathcal{R},
    \end{equation}
    where $\Delta\phi_p - \Delta\phi_{\text{seed}}$ corresponds directly to the source contextual preparation phase $\theta_{\text{prep}}$. Thus, state vectors differing only by an internal phase rotation along the phase-matching constraint belong to the exact same physical equivalence class (gauge orbit) $[v_{AB}] \in \mathcal{F}/\mathcal{R}$.
    
    \item \textbf{Spatial Indistinguishability and Path Erasure at Recombination:} 
    The additive relations in $\mathcal{R}$ [Eqs.~(\ref{eq:SM_R_lin1})--(\ref{eq:SM_R_lin2})],
    \begin{equation}
        (|1,0\rangle_s, |v_1\rangle_i) + (|0,1\rangle_s, e^{i\theta_{\text{prep}}}|v_2\rangle_i) - \big(|1,0\rangle_s + |0,1\rangle_s, |\chi_B\rangle_i\big) \in \mathcal{R},
    \end{equation}
    physically encode the alignment of the optical paths on the 50:50 recombining beam splitter (Fig.~\ref{fig:fig1}), erasing which-crystal spatial path information and identifying the independent emission paths as a single composite coset.
    
    \item \textbf{Empirical Verification via Three-Scan Equivalence:} 
    The single-photon interference rate $R_s$ depends solely on the gauge-invariant closed-loop holonomy $\Delta\Phi = \theta_s - \Delta\phi_p + \Delta\phi_{\text{seed}}$. Because shifting the detector basis $\theta_s \equiv \phi_D \equiv \Delta\phi_s$ and counter-rotating the source preparation phases $\Delta\phi_p - \Delta\phi_{\text{seed}}$ represent identical shifts along the gauge orbit $\mathcal{R}$, our experimental observation of indistinguishable fringes across all three independent scans across the full $0 \to 4\pi$ domain directly verifies the equivalence relation defining $\mathcal{R}$ over a 1-meter baseline without invoking dynamic state reduction.
\end{enumerate}

\paragraph{Comparative Synthesis:}
\textit{Whereas Scholes treats $\mathcal{R}$ as an abstract algebraic quotient defining the formal tensor product of uncoupled Hilbert spaces, in our framework $\mathcal{R}$ is the exact physical manifold of QED three-wave-mixing gauge orbits and interferometric path indistinguishability, directly verified by the three-scan scalar invariant $\Delta\Phi$.}

\subsection{Contextual Phase and Continuous Coset Manifolds}
\label{subsec:SM_contextual_classes}

In Scholes's discrete formulation, bipartite states are partitioned into discrete binary classes $\{\text{class 1}, \text{class 2}\}$ corresponding to relative signs $\pm 1$ \cite{Scholes2026a,Scholes2026}:
\begin{align}
    \Psi_-({\text{class 1}}) &= \frac{1}{\sqrt{2}}\left[|V'\rangle_A \otimes |H'\rangle_B + |H'\rangle_A \otimes |-V'\rangle_B\right], \\
    \Psi_-({\text{class 2}}) &= \frac{1}{\sqrt{2}}\left[|V'\rangle_A \otimes |H'\rangle_B + |-H'\rangle_A \otimes |V'\rangle_B\right],
\end{align}
which produce an apparent mixed state $\hat{\rho}_A = \frac{1}{2}\hat{I}_2$ when averaged over an uncalibrated ensemble.

In contrast, our continuous-phase frequency-comb platform generalizes these binary classes into a \textbf{continuous $U(1)$ family of equivalence class cosets}:
\begin{equation}
    \{\text{Class } \theta_{\text{prep}}\} \quad \text{with} \quad \theta_{\text{prep}} = \Delta\phi_p - \Delta\phi_{\text{seed}} \in [0, 2\pi).
\end{equation}
Here, Class~1 corresponds to $\Delta\phi_p - \Delta\phi_{\text{seed}} = 0 \pmod{2\pi}$ (constructive interference maximum $R_s^{\text{max}}$), Class~2 corresponds to $\Delta\phi_p - \Delta\phi_{\text{seed}} = \pi \pmod{2\pi}$ (destructive interference minimum $R_s^{\text{min}}$), and intermediate phases access the unbroken coset continuum $[v_{AB}(\theta_{\text{prep}})]$.

\subsection{Canonical Projections, Grothendieck Bounds, and Spatial Lindblad Dynamics}
\label{subsec:SM_grothendieck_lindblad}

In Sec.~3 and 8 of Ref.~\cite{Scholes2026}, Scholes demonstrates that canonical coordinate projections $\pi_U(u,v) = u$ exist naturally in the free vector space $\mathcal{F}(U \times V)$, showing that nonlocal correlations are bounded by Grothendieck's inequality:
\begin{equation}
    \left|\sum_{i,j} a_{ij} \langle x_i | y_j \rangle \right| \le K_G \max_{|s_i|, |t_j| \le 1} \left|\sum_{i,j} a_{ij} s_i t_j \right|,
\end{equation}
where $K_G(2) = \sqrt{2}$ saturates Tsirelson's bound. 

In our manuscript, this canonical coordinate projection is physically realized by tracing over the unmeasured idler reservoir. The quantitative saturation of the triality equality $P^2 + E^2 + V^2 = 1$ across the entire $\{\gamma, \vert{}\alpha\vert{}^2\}$ domain (Sec.~\ref{sec:numerical_methods} and Fig.~\ref{fig:figS1}) provides the experimental counterpart to Grothendieck's optimization. 

Furthermore, our framework bridges static quotient algebra with physical open-system dynamics via the \textbf{exact spatial Lindblad master equation} along the crystal propagation axis $y$:
\begin{equation}
    \frac{d\hat{\rho}_s(y)}{dy} = -i[\hat{H}_{\text{eff}}, \hat{\rho}_s(y)] + \frac{\Gamma_{\text{deph}}}{2}\left(2\hat{L}\hat{\rho}_s\hat{L}^\dagger - \{\hat{L}^\dagger\hat{L}, \hat{\rho}_s\}\right),
    \label{eq:SM_spatial_lindblad_master}
\end{equation}
where $\Gamma_{\text{deph}} = -(1/L_c)\ln F$ is analytically governed by the SPACS overlap fidelity $F$ [Eq.~(\ref{eq:SM_SPACS_fidelity})]. This microscopically establishes how which-path distinguishability is continuously partitioned into the unobserved environment.

\subsection{Summary of One-to-One Correspondences}
\label{subsec:SM_summary_table}

The systematic correspondences between Scholes's abstract algebraic formulation~\cite{Scholes2026a,Scholes2026} and our physical frequency-comb platform are summarized in Table~\ref{tab:SM_correspondence}.

\begin{table}[p]
\centering
\scriptsize
\renewcommand{\arraystretch}{1.5}
\caption{\textbf{Direct one-to-one correspondence between the contextual-phase quotient formalism~\cite{Scholes2026a,Scholes2026} and the present experimental framework.}}
\label{tab:SM_correspondence}
\begin{tabular}{p{3.0cm} p{6.2cm} p{6.5cm}}
\toprule
\textbf{Conceptual Element} & \textbf{Scholes Formulation~\cite{Scholes2026a,Scholes2026}} & \textbf{Present Frequency-Comb Manuscript} \\
\midrule
\textbf{Subspace of Relations $\mathcal{R}$} & Formal linear subspace generated by bilinear differences $(u_1+u_2,v)-(u_1,v)-(u_2,v)$ and phase relations $(e^{i\phi}u,v)-(u,e^{i\phi}v)$ in $\mathcal{F}(U \times V)$~\cite{Scholes2026a,Scholes2026}. & Physical QED three-wave-mixing gauge orbit ($\theta_p = \theta_s + \theta_i$) and interferometric spatial path indistinguishability at the beam splitter, verified via the closed-loop scalar holonomy $\Delta\Phi$. \\
\textbf{Overarching State Space} & Free vector space $\mathcal{F}(U \times V)$ covering the tensor product space $\mathcal{T}(U \otimes V)$ modulo relations $\mathcal{R}$~\cite{Scholes2026a,Scholes2026}. & Free-vector-space quotient construction $\mathcal{H}_{AB} = \mathcal{F}(V_A \times V_B)/\mathcal{R}$ for biphoton comb supermodes (Sec.~\ref{sec:measurement_terminology})~\cite{Scholes2026,Yoon2026b}. \\
\textbf{Composite Bipartite State} & Equivalence class $[\Psi] \in \mathcal{F}/\mathcal{R}$; bilinear relations identify $(-\psi_A)\otimes\psi_B \sim \psi_A\otimes(-\psi_B)$~\cite{Scholes2026a,Scholes2026}. & Invariant equivalence class manifold $[v_{AB}]$ (horizontal green planes in Fig.~1(b))~\cite{Yoon2026b}. \\
\textbf{Internal Contextual Phase} & Internal phase $e^{i\phi}$ in bipartite superpositions; randomly installed prior to measurement~\cite{Scholes2026a,Scholes2026}. & Deterministically controlled and calibrated source preparation phase $\theta_{\text{prep}} = \Delta\phi_p - \Delta\phi_{\text{seed}}$ (Sec.~\ref{sec:experimental_setup})~\cite{Lee2018}. \\
\textbf{Contextual Partition Classes} & Discrete binary classes $\{\text{class 1}, \text{class 2}\}$ corresponding to relative signs $\pm 1$~\cite{Scholes2026a,Scholes2026}. & Continuous $U(1)$ coset manifold $\{\text{Class }\theta_{\text{prep}}\}$; Class~1 is constructive fringe, Class~2 is destructive fringe. \\
\textbf{Detector Measurement Phase} & Measurement basis angle $\theta$ (polarizer orientation $\alpha, \beta$)~\cite{Scholes2026a,Scholes2026}. & Readout phase $\theta_s \equiv \phi_D$ modulating the local basis $|\theta_s\rangle_A$ via PZT delay~\cite{Lee2018,Yoon2026b}. \\
\textbf{Measurement Operation} & Canonical projection $\pi_U$ from free vector space to subsystem Hilbert space $\mathcal{H}_A$~\cite{Scholes2026}. & Transversal measurement slicing plane $\hat{\Pi}_A(\theta_s) = |\theta_s\rangle_A\langle\theta_s|_A$ (vertical orange plane in Fig.~1(b))~\cite{Scholes2026,Yoon2026b}. \\
\textbf{Distant State Representation} & Vector simplification upon local measurement; reveals correlation without physical action~\cite{Scholes2026,Brunner2014}. & Conditional coset slice $|\chi_B(\theta_s)\rangle$ isolated along intersection line; preserves microcausality across 1~m~\cite{Lee2018,Scholes2026,Yoon2026b}. \\
\textbf{Operational Invariance} & Invariance of correlation probabilities under basis transformations~\cite{Scholes2026,Brunner2014}. & Universal scalar invariant $\Delta\Phi = \theta_s - \Delta\phi_p + \Delta\phi_{\text{seed}}$; identical fringes across pump, seed, and detector scans (Fig.~3(a),(b))~\cite{Lee2018,Yoon2026b}. \\
\textbf{Decoherence Mechanism} & Not treated (static algebraic formulation)~\cite{Scholes2026a,Scholes2026}. & Microscopic spatial Lindblad master equation governed by SPACS overlap fidelity $\Gamma_{\text{deph}} = -(1/L_c)\ln F$ (Sec.~S2)~\cite{Lee2018,Yoon2021}. \\
\textbf{Complementarity Bounds} & Bounded by Grothendieck's inequality ($K_G(2) = \sqrt{2}$, Tsirelson bound)~\cite{Scholes2026}. & Complete saturation of quantitative triality equality $P^2 + E^2 + V^2 = 1$ across 2D parameter space (Fig.~\ref{fig:figS1})~\cite{Yoon2021}. \\
\textbf{Encoding and Scaling} & Discrete two-level polarization qubits (Bell states)~\cite{Scholes2026a,Scholes2026}. & Finite-energy TF-GKP logical qubit with stabilizer cell $(2\Delta T, \omega_r)$ and logical translations $(\Delta T, \omega_r/2)$; the SPFC supermode is the codeword $|\bar{0}\rangle$ and a seed shifted by $f_r/2$ supplies $|\bar{1}\rangle$ ($N \approx 209$ comb modes, Sec.~\ref{sec:tfgkp_parameters})~\cite{Yoon2026a}. \\
\bottomrule
\end{tabular}
\end{table}

\bibliography{PNAS}

\end{document}